\documentclass[]{aa}
\usepackage[colorlinks=true,linkcolor=blue,citecolor=blue]{hyperref}
\usepackage{lipsum,multicol}
\usepackage{graphicx}

\def \ms{m\,s$^{-1}$ }

\def \mjup{M$_\mathrm{Jup}$}

\def \Gaia{\textit{{Gaia} }}
\def \Hipparcos{\textsc{\large{Hipparcos}} }
\def \HipGaia{\textsc{\large{Hipparcos}}-\textit{Gaia} }
\bibpunct{(}{)}{;}{a}{}{,}
\usepackage{txfonts}
\usepackage{soul}
\usepackage{pdflscape}
\usepackage{capt-of}
\usepackage{placeins}
\DeclareMathAlphabet{\mathpzc}{OT1}{pzc}{m}{it}

\begin{document}

   \title{True masses using RV data with Hipparcos and Gaia Astrometry}

   \subtitle{}

   \author{G. Piccinini\inst{1,2}
           \and A. Petralia\inst{1} \and A. Sozzetti\inst{3} \and S. Benatti\inst{1} \and D. Gandolfi\inst{4} \and G. Micela\inst{1}
          }

   \institute{INAF -- Osservatorio Astronomico di Palermo, Piazza del Parlamento 1, I-90134, Palermo, Italy\\ 
   \email{giulia.piccinini@inaf.it}
   \and
            Dipartimento di Fisica e Chimica, Università degli Studi di Palermo, Via Archirafi 36, I-90123, Palermo, Italy
             \and
              INAF -- Osservatorio Astrofisico di Torino, Strada Osservatorio 20, I-10025, Pino Torinese (TO), Italy \and
              Dipartimento di Fisica, Università degli Studi di Torino, Via Pietro Giuria 1, I-10125, Torino, Italy
             }

   \date{Received 4 August 2025; Accepted 13 January 2026}

  \abstract
  {Long-period companions are detected and characterized thanks to long-baseline radial velocity surveys. Combining Doppler time-series with astrometry, and in particular with proper motion anomalies technique, it is possible to put strong constraints on their orbital inclination and true mass.}
   {This work aims to present a model that combines Hipparcos and Gaia astrometric data with radial velocity measurements to constrain the orbital inclinations and true masses of long-period companions. Additionally, we re-analyse a small sample of targets that have not yet been studied using this combined approach.
   }
   {This research leverages the simultaneous modelling of proper motion anomalies and radial velocities, in conjunction with an analysis of the sensitivity curve. This approach serves not only as a verification of the parameters but also as a means to acquire valuable insights into planetary systems.
   }
   {The new analyses reveal that some of the targets classified as brown dwarfs or small-mass stars have a planetary nature. HD 5388 b and HD 6718 b are likely planets, with masses of $3.2_{-0.2}^{+0.3}$ $\mathrm{M_{Jup}}$ and $2.4_{-0.2}^{+0.3}$ $\mathrm{M_{Jup}}$, respectively. HD 141937 b is likely a planet, but the current dataset does not allow us to firmly constrain its true mass. HD 16760 b belongs to the brown dwarf regime and it has a probable second companion. 30 Ari B b falls within the stellar regime, but the presence of an additional stellar companion could compromise the reliability of the final results. For HD 148427 b, HD 96127 b and HIP 65891 b we determined a range for the orbital inclinations.
   }
  {}


   \maketitle

\section{Introduction}
\label{sec:introduction}
Mass is a fundamental property for the characterization of an orbiting companion, a critical parameter for assessing its nature. A neglected mass determination can lead to ambiguities. Objects with sufficient mass to initiate thermonuclear hydrogen fusion in their cores are defined as stars (M $\gtrsim$ 80 $\mathrm{M_{Jup}}$). Substellar objects are classified as (exo)planets, with masses up to $\sim$ 13 $\mathrm{M_{Jup}}$ (the critical mass for deuterium fusion), and brown dwarfs (BDs), which occupy the intermediate mass range between planets and stars \citep{Burrows2001,hatzes2005}.
Exoplanet population exhibits remarkable diversity, such as super-Earths, Neptunes, hot and warm gas giant, cold and cool gas giant \citep{cloutierdemograhics}. However, for several hundred planets, the true mass remain undetermined, with only a known lower limit. 

Long-period gas giant planets (\textit{a} = 1 -- 10 au, 0.3 $\leq$ $\mathrm{M_{c}}$ $\leq$ 13 $\mathrm{M_{Jup}}$) may influence the formation of smaller planets such as sub-Neptunes and super-Earths.
According to the inward migration models, giant planets can act as barriers that inhibit the inward migration of smaller planets \citep{izidoro2015gas}, or if the giant planets form from cores farther out in the disk, smaller planets originating near the snowline may still migrate inward freely \citep{2015A&A...582A.112B}. In contrast, the in-situ scenarios consider the influence of giant planets either by blocking inward pebble flux \citep{2019A&A...627A..83L} or by involving massive disks that concurrent formation of both planetary populations \citep{2013MNRAS.431.3444C}.

Recently, \cite{Pinamonti2023CJ} suggested that around late-type dwarf stars, long-period planets might instead promote the formation of inner sub-Neptunes.
A comprehensive understanding of planet formation and evolution also depends on accurately determining the occurrence rate of gas giant exoplanets. This requires expanding and refining the sample of known gas giants with well-constrained masses. Most occurrence estimates come from radial velocity (hereafter RV) measurements, thanks to the long observational baselines. However, being a one dimensional technique, it cannot directly determine the true mass of a companion $\mathrm{M_{c}}$, providing only the lower limit expressed as $\mathrm{M_{c}}\sin{i}$, where \textit{i} is the orbital inclination, with respect to the observer, and so they may include contaminants such as brown dwarfs or very low-mass stars. As a result, the true occurrence rate of giant planets could be overestimated.
For example, using HARPS-N/TNG \citep{HARPSN} spectrograph and archival Doppler measurements, \cite{bonomo2023cold} found that the occurrence rate of gas giant planets is $9.3^{+7.7}_{-2.9}\%$, in system with FGK stars and known to host inner small planets. This agrees with the anti-correlation between the presence of small planets and gas giant planets predicted by \cite{izidoro2015gas}. More recently, \citet{bonomo2025cjsmpl} extended the analysis by accounting for stellar metallicity for a sample of solar-type stars, finding an occurrence rate of $11.1^{+2.5}_{-1.8}\%$. However, since they considered the minimum mass, the occurrence rate might decrease if the $\mathrm{M_{c}}\sin{i}$ of the considered giant planets turned into true mass measurements above the 13 $\mathrm{M_{Jup}}$ upper limits.

It is necessary to combine the RV data with another method, in order to overcome the $\sin{i}$ ambiguity.
The planet’s mass can be determined by measuring its gravitational influence on the host star and one technique that take advantage of this effect is astrometry. It involves the precise measurement and analysis of the positions and motions of celestial bodies on the sky compared to other stars \citep{malbet2018astrometry}. 
This technique can confirm the nature of planets candidate, in combination with RV method. In this way, all the orbital parameters can be assessed (orbital period \textsc{P}, time of the pericentre passage $\mathrm{T_{p}}$, argument of the pericentre \textsc{$\omega$}, semi-amplitude \textsc{K}, eccentricity e, orbital inclination \textit{i} and longitude of the ascending node \textsc{$\Omega$}), including the precise mass of the companion, $\mathrm{M_{c}}$. 
Over the past 35 years, astrometric measurements have advanced significantly thanks to positional observations made by satellites operating above Earth's atmosphere \citep{perryman2012history}. European Space Agency (ESA)'s space missions \Hipparcos \citep[High Precision Parallax Collecting Satellite, 1989-1993;][]{perryman1997hipparcos, vanLeeuwen1997}, and \Gaia \citep[Global Astrometric Interferometer for Astrophysics;][]{prusti2016gaia}, launched in December 2013 and completed its mission in January 2025, have significantly enhanced the precision and capabilities of astrometric techniques, providing a substantial boost to the field. 
Taking advantage of the temporal separation between the two missions, several authors have developed models to constrain the true masses of stellar companions by combining absolute and relative astrometry with RV measurements \citep[e.g.,][]{Tokovinin1992, orbitize, damasso2020precise, xuan2020evidence, orvara, kiefer2021determining, octofitter}.
Unlike the RV method, astrometry is well-suited for detecting planets across a wider range of periods, as the astrometric signal increases with the semi-major axis of a planet. 

In this study, we combined RV technique with astrometry to determine the true masses of a selected sample of objects. To validate the performance of our model, we first tested it on three known cases from literature. 
We then applied our model to the targets presented in \cite{kiefer2021determining}. They use \Gaia DR1 \citep{GDR12016} to simulate astrometric noise in order to constrain the orbital inclinations and derive true masses. Astrometric excess noise is a measure of the residual scatter around the five-parameter astrometric solution as computed by the \Gaia reduction pipeline. In GDR1, all sources are treated as single stars, and any astrometric signal caused by binary motion is ignored. The authors used the Gaia Astrometric noise Simulation To derived Orbit iNclination \citep[\texttt{GASTON};][]{kiefer2019detection} tool to simulate \Gaia photocentre measurements along the RV-constrained orbits, testing various orbital inclinations with respect to the plane of the sky. Each simulated inclination produced a different astrometric excess noise, which was then compared to the actual GDR1 measurement to constrain the inclination.
\cite{kiefer2021determining} applied their method to 755 planets, orbiting 658 stars. Among these companions, for eight systems (HD 5388, HD 6718, HD 16760, 30 Ari B, HD 141937, HD 148427, HIP 65891) that exhibit significant astrometric excess noise, the authors were able to determine orbital inclinations and thus the true companion masses, classifying them as brown dwarfs, low-mass stars, or planets. For HD 96127, although \texttt{GASTON} did not reach convergence, the derived mass lies in the stellar regime, yet it remains compatible with the planetary domain within 1$\sigma$.

This work aims to independently verify the results of \cite{kiefer2021determining}, which focus on excess noise, by presenting the first analysis of these systems using a joint spectroscopic and astrometric modelling technique.
The article is organized as follows: in Section \ref{sec:methods} we describe the method that we used in our analysis; in Section \ref{sec:validation} we discuss the validation of our code by applying it to three targets; in Section \ref{sec:Results} we show the results of our study using the sample of objects reported in \cite{kiefer2021determining}.

\section{Methods}
\label{sec:methods}
\subsection{Proper motion anomaly}
\label{sec:PMA}
The astrometric influence of a companion in long-period orbits can be investigated through proper motion anomaly (hereafter PMA), which quantifies variations in a star's measured proper motion over time due to orbital acceleration \citep{2018ApJS..239...31B,kervella2019stellar,brandt2021hipparcos}. 
The presence of a hidden companion can thus be detected if a star exhibits considerable different proper motions when comparing observations taken with significant temporal separation. The data acquired by \Hipparcos and \Gaia satellites can satisfy this requirement since the temporal gap between the two missions is more than two decades. In this way, by defining a long-term proper motion vector for each star and comparing it with the short-term proper motion vectors from \Hipparcos or \textit{Gaia}, one can study the properties of an orbiting companion through its influence on the primary star's motion. The PMA technique analyses the differences between the system centre of mass proper motion and the orbital proper motion.
PMA is most sensitive to companions with orbital periods ranging from approximately 3 year (since the proper motion measurements are not instantaneously measured quantities, they are affected by ``smearing effect'', as they represent time-averaged values over the observing time window of each mission) to a few times 25 years, the baseline between \Hipparcos and \textit{Gaia}. For shorter periods, the \Gaia proper motion averages over multiple orbits and thus loses its sensitivity; for longer periods, the orbital motion is absorbed into the longer-term \HipGaia proper motion. Additionally, PMA sensitivity scales inversely with stellar distance, making it most effective for the nearby stars \citep{el2024gaia}.

Our model performs a simultaneous analysis combining absolute astrometry—in particular using the PMA technique—and RV data. The core structure of our model is based on the equations presented in \cite{xuan2020evidence}. 

In the literature, the two main astrometric acceleration catalogues are those by \cite{kervella2019stellar} and \cite{2018ApJS..239...31B} (\HipGaia Catalog of Accelerations, HGCA), with a later updated version published by \cite{brandt2021hipparcos}. 
In HGCA, the authors did a cross-calibrated catalogue of \textsc{\large{Hipparcos}}, \Gaia DR2 \citep{GDR22018} and DR3 (or EDR3) \citep{GEDR3} (hereafter, GDR2 and GDR3 respectively)
astrometry to enable their use in measuring changes in proper motion, i.e., accelerations in the plane of the sky, that places both on a common reference frame with calibrated uncertainties.
Our analysis is based on the catalogues by \cite{kervella2019stellar} for GDR2 and \cite{kervella2022stellar} for GDR3, since we also use the sensitivity curve (see Section \ref{SensitivityCurve}), the theoretical tool that defines the expected companion mass as a function of the semi-major axis, described in Kervella et al. (\citeyear{kervella2019stellar}). In these catalogues, for each star are reported the PMA components ($\Delta\mu_{\alpha}$, $\Delta\mu_{\delta}$), i.e. the projected velocity vector of the photocentre around the barycentre for \Hipparcos at the epoch 1991.25, for GDR2 at the epoch 2015.5 and for GDR3 at the epoch 2016.0.
Since the proper motions used to compute the PMAs are an average over the observing time window of the missions, we retrieved the observational epochs from ``The Hipparcos and Tycho Catalogues'' \citep{hipparcos} and the ``Gaia Observation Forecast Tool'' (\href{https://gaia.esac.esa.int/gost/index.jsp}{\texttt{GOST}}\footnote{https://gaia.esac.esa.int/gost/index.jsp}) and averaged them over their respective observational windows. When necessary, the epochs were adjusted for the differences in the Earth's position relative to the Solar System's barycentre, and all the dates were converted to Barycentric Julian Date in Barycentric Dynamical Time ($\mathrm{BJD_{TDB}}$).

To determine the precise mass of our targets, we considered as parameters the five spectroscopic ones (\textsc{P}, $\mathrm{T_{p}}$, $\sqrt{e}\sin{\omega}$, $\sqrt{e}\cos{\omega}$, \textsc{K}), \textsc{$\Omega$}, $\cos{i}$, the RV offset $\gamma$ and the jitter $\sigma$. The RV offset and the jitter terms are derived for every dataset and instrument considered for each target. The jitter term accounts for additional noise not included in the RVs errors and it is added in quadrature to them. 

For the RV component of the model, we fitted the orbital parameters using a Keplerian function. Specifically, we used the \texttt{radvel} package \citep{fulton2018radvel}\footnote{\textsc{radvel.kepler.rv$\_$drive, radvel.orbit.timetrans$\_$to$\_$timeperi}} to assess the spectroscopic values. For the astrometric component, we used the spectroscopic parameters together with $\cos{i}$ and $\Omega$ to calculate the Thiele–Innes constants. From these values, we derived the predicted positions and proper motions. To account for the smearing effect, we averaged the instantaneous modelled orbital positions and proper motions. To jointly integrate the PMA and RV models, we then calculated the log-likelihood function.

We follow the expression described by \cite{xuan2020evidence}, as detailed below:
\begin{equation}
    \mathrm{log}\mathcal{L} = -\frac{1}{2}\Big(\chi_{\mathrm{PMA}}^{2} + \chi_{\mathrm{RV}}^{2}\Big).
\end{equation}
The PMA part is described as
\begin{equation}
    \chi_{\mathrm{PMA}}^{2} = \sum_{i}^{H,G}\Bigg[\Bigg(\frac{\mathcal{M}[\Delta\mu_{i,\alpha}] - \Delta\mu_{i,\alpha}}{\sigma(\Delta\mu_{i,\alpha})}\Bigg)^{2} + \Bigg(\frac{\mathcal{M}[\Delta\mu_{i,\delta}] - \Delta\mu_{i,\delta}}{\sigma(\Delta\mu_{i,\delta})}\Bigg)^{2}\Bigg].
\end{equation}
We considered separately the two PMA coordinates ($\Delta\mu_{\alpha}$ and $\Delta\mu_{\delta}$) and the two catalogues (\Hipparcos and \textit{Gaia}). $\mathcal{M}[\Delta\mu_{\alpha,\delta}]$ are the results obtained from the model, $\Delta\mu_{\alpha,\delta}$ and $\sigma(\Delta\mu_{\alpha,\delta})$ are the values and the errors from Kervella's catalogues.

The RV part for one instrument is described as 
\begin{equation}
    \chi_{\mathrm{RV}}^{2} = \sum_{j}\Bigg[\frac{\Big(\mathcal{M}[\mathrm{RV}_j] - \mathrm{RV}_{j} + \gamma\Big)^{2}}{\sigma(\mathrm{RV}_j)^2 + \sigma^2} + \mathrm{log}\text{ }2\pi(\sigma(\mathrm{RV}_j)^2 + \sigma^2)\Bigg],
\end{equation}
where $\mathcal{M}[\mathrm{RV}_j]$ is the result from the model obtained by using \texttt{radvel} at time \textit{j}; $\mathrm{RV}_j$ and $\sigma(\mathrm{RV}_j)$ are the data and the uncertainties at time \textit{j}. $\gamma$ and $\sigma$ are the RV offset and the jitter term. 

We performed a Differential Evolution Markov Chain Monte Carlo (DE-MCMC) for the simultaneous analysis \citep{ford2005quantifying,braak2006markov}. Since, for most of the systems, this is the first time that a simultaneous analysis of RV and PMA has been carried out, we did not start our chains with strong priors based on results from the literature; instead, we chose to use informative uniform priors with a wide range around the known values. For $\cos{i}$ and $\Omega$, we explored the full allowed range, considering prograde (\textit{i} $<$ 90°) and retrograde (\textit{i} $>$ 90°) solutions separately. We used the \texttt{emcee} package \citep{emcee} and first performed a burn-in phase corresponding to 10\% of the total number of steps, followed by 500,000 steps for each target. We employed 96 random walkers for the analysis.

The medians of the posterior distributions are taken as the parameter values, while the 1$\sigma$ uncertainties are derived by evaluating the $\pm$ 34\% intervals of the posteriors with respect the median. 
We checked that all the chains exceeded 50 times the integrated autocorrelation time for each parameter. As additional check, we assessed the convergence of the chains using the Gelman–Rubin statistic and the $\mathrm{\widehat{R}}$ parameter \citep{GelmanRubin,Ford2006}. In particular, following \cite{Rhat_new}, all the results meet the criterion of $\mathrm{\widehat{R}}$ $< 1.001$. Finally, thanks to the simultaneous analysis, we derived precise mass values by evaluating all the parameters required to compute the true companion mass $\mathrm{M_{c}}$ from the expression for \textsc{K}, as presented here

\begin{equation}
    K = \bigg(\frac{2\pi G}{P}\bigg)^{1/3} \frac{M_{c}\sin i}{M_{\star}^{2/3}} \frac{1}{(1-e^{2})^{1/2}};
\end{equation}
where G is the gravitational constant and $\mathrm{M_{\star}}$ is the star mass. We then derived the mass posterior distributions and, similarly to the other parameters, adopted the medians as estimates of the true values, and the 16\% and 84\% quantiles as the uncertainties.

\subsection{Validation of the results: Sensitivity Curve and RUWE}
\label{SensitivityCurve}
The sensitivity curve is the final check we perform to validate the obtained results. The theoretical framework of the sensitivity curve is described in \cite{kervella2019stellar}. Here, we explicit the equation:
\begin{equation}
    M_{c}(r) = \frac{\sqrt{r}}{\gamma}\sqrt{\frac{M_{\star}}{G}}\frac{\Delta\nu_{T}}{\eta\xi},
\end{equation}
where the parameter $\gamma$ quantifies the relative sensitivity variation of the PMA to orbiting companions, due to the observing window smearing effect; $\Delta\nu_{T}$ is the tangential velocity anomaly; $\eta$ is a corrector factor for the effects of orbital inclination and eccentricity; $\xi$ minimizes biases that may arise when the orbital period is longer than the time baseline between \Hipparcos and \Gaia (i.e., 24.25 years for GDR2). The mass of the primary star, $\mathrm{M_{\star}}$, and the gravitational constant, G, are the last elements in the equation.
We refer to \cite{kervella2019stellar} for a detailed explanation of the equation.

Thanks to this tool, we are capable of finding possible combinations of the semi-major axis and the companion's mass. Specifically, we exploit the astrometric signal of the stellar companion in the form of $\Delta\nu_{T}$.
The sensitivity curve exhibits several characteristic ``spikes'' at distance \textit{r}. Since the PMA represents a time-averaged velocity vector of the star over the observational window ($\delta_{t}$), the spikes arise because the PMA signal becomes null when the orbital period, equal to the observational time window, is divided by an integer of $\delta_{t}$ of the respective catalogues ($\delta t_{\mathrm{H}}$ = 1227 days, $\delta t_{\mathrm{GDR2}}$ = 668 days, $\delta t_{\mathrm{GDR3}}$ = 1038 days). 
As a result, when \textsc{P} is significantly shorter than $\delta$t, the PMA signal is diminished due to temporal smearing \citep{kervella2019stellar}.
This effect leads to a loss of sensitivity, resulting in the non-detection of companions independently of their mass \citep{kervella2022stellar}. 
To mitigate this issue, we adopt data from GDR2 or GDR3 when the orbital period is comparable to the observational time window, taking into account the proper $\Delta\nu_{T}$. 

A limitation of the sensitivity curve occurs when one of the most comprehensive observational dataset (\Hipparcos or GDR3) does not cover the periastron passage of the companion, but instead observed the apastron. In such cases, the tangential velocity anomaly of the host star is significantly reduced due to the slower orbital velocity at apastron. This effect highlights an intrinsic constrain of the PMA analysis technique, which assumes a circular orbit for the companion and accounts for the uncertainty in orbital inclination through statistical methods. However, the influence of orbital eccentricity on the overall analysis is generally limited, as it primarily affects individual measurements rather than the long-term proper motion vector \citep{kervella2022stellar}.

We also considered the renormalised unit weight error \citep[RUWE;][]{lindegren2018gaia}, as a check on the results obtained. RUWE indicates the quality of the 5-parameter solution provided by \textit{Gaia}.
Typically, a RUWE value of 1.4 is used as a threshold to distinguish between good and poor solutions, though this has been the subject of various studies.
\cite{StassunTorres} found that when RUWE is in the range 1.0-1.4, its values are strongly correlated with photocentre motion and sensitive to unresolved companion and can serve as quantitative predictor of motion. 
For GDR3, \cite{Penoyre22a} recommended a threshold of 1.25 to identify potential binary, specifically for nearby targets (distance $<$ 100 pc). In the absence of detailed time-series astrometry, RUWE remains the most reliable metric for detecting companions from astrometric data alone \citep{HD141937Wallace}.

\section{Validation}
\label{sec:validation}
We applied our model on three objects to test its accuracy. We considered the super-Jupiter GJ 463 b \citep{endl2022jupiter}, comparing our results with the ones obtained by \citep{sozzetti2023dynamical}; the brown dwarf $\pi$ Men b \citep{damasso2020precise}; the long-period super-Jupiter HD 222237 b \citep{xiao2024hd}. The literature values for the three cases were derived using a combination of RV and astrometric data. For GJ 463 b and $\pi$ Men b analogues methodologies were applied utilizing GDR3 and GDR2 datasets, respectively. Notably, $\pi$ Men b has been the subject of extensive astrometric investigations by multiple studies. HD 222237 b constitutes a particularly valuable test case due to its long orbital period of nearly 40 years, providing a unique opportunity to assess the model's performance on very long-period companions, as well as for the characterization technique employed. In the following section, we will describe the different approaches used. In Table \ref{tab:allvalidation}, the three planetary systems' properties and the derived planetary parameters are reported.

\begin{table*}
    \centering
    \small
   \caption{Properties of the three planetary systems used for the validation of the code.}
    \renewcommand\arraystretch{1.2}
    \begin{tabular}{l|cccccc}
    \hline
    \hline
       Parameters & GJ 463 $^{(a,b,c,d)}$ && $\pi$ Men $^{(a,b,c,e)}$ && HD 222237 $^{(b,f,g,h)}$ & \\ 
    \hline
        RA [J2000] & 12:23:00.16 && 05:37:09.89 && 23:39:37.39& \\
        Dec [J2000] & +64:01:50.96 && -80:28:08.83 && -72:43:19.76 & \\
        Parallax $\pi$ [mas] & 54.45 $\pm$ 0.03 && 54.68 $\pm$ 0.04 && 87.37 $\pm$ 0.02 & \\
        Distance \textit{d} [pc] & 18.37 $\pm$ 0.01 && 18.27 $\pm$ 0.02 && 11.445 $\pm$ 0.003 &\\
        $m_{V}$ & 13.48 $\pm$ 0.01 && 5.65 $\pm$ 0.01 && 7.09 & \\
        Radius $\mathrm{R_{\star}}$ [R$_{\odot}$] & 0.49 $\pm$ 0.02 && 1.10 $\pm$ 0.01 && 0.71 $\pm$ 0.06 & \\
        Mass $\mathrm{M_{\star}}$ [M$_{\odot}$] & 0.49 $\pm$ 0.02 && 1.02 $\pm$ 0.03 && 0.76 $\pm$ 0.09 &\\
        Effective temperature $\mathrm{T_{eff}}$ [K] & 3540 $\pm$ 41 && 5870 $\pm$ 50 && 4751 $\pm$ 139 &\\
        Spectral type & M3V && G0V && K3V &\\
        \hline
        $\Delta\mu_{\alpha}$ [mas yr$^{-1}$] (\textsc{Hipparcos}) & 4.330 $\pm$ 2.001 && 1.057 $\pm$ 0.240 && -0.896 $\pm$ 0.071&\\
        $\Delta\mu_{\delta}$ [mas yr$^{-1}$] (\textsc{Hipparcos}) & -0.703 $\pm$ 2.141 && 0.787 $\pm$ 0.260 && 0.055 $\pm$ 0.420&\\
        $\Delta\mu_{\alpha}$ [mas yr$^{-1}$] (\textit{Gaia} DR3) & -0.641 $\pm$ 0.075 && 0.165 $\pm$ 0.047 && 0.892 $\pm$ 0.026&\\
        $\Delta\mu_{\delta}$ [mas yr$^{-1}$] (\textit{Gaia} DR3) & 0.311 $\pm$ 0.058 && -0.586 $\pm$ 0.060 && -0.232 $\pm$ 0.027&\\
        $\Delta\nu_{T}$ [m s$^{-1}$] (\textit{Gaia} DR3) & 61.94 $\pm$ 8.24 && 52.79 $\pm$ 6.65 && 49.99 $\pm$ 2.02 &\\
        RUWE & 1.407 && 0.814 && 1.049\\
        \hline
        &Endl (\citeyear{endl2022jupiter})&This Work& Damasso (\citeyear{damasso2020precise})&This Work&Xiao (\citeyear{xiao2024hd})&This Work\\
        \textsc{K} [ms$^{-1}$] & 33 $\pm$ 3 &$33.7_{-3.0}^{+3.3}$ & 196.1 $\pm$ 0.7 &195.3 $\pm$ 0.8 & $46.57_{-0.82}^{+0.83}$&47.2 $\pm$ 0.9 \\
        \textsc{P} [d] & $3448_{-88}^{+110}$ &$3453.5_{-83.0}^{+92.9}$ & 2088.8 $\pm$ 0.4 &2089.0 $\pm$ 0.4& $13262_{-1165}^{+1319}$&$13997_{-1397}^{+1514}$ \\
        $\mathrm{T_{p}}$ [BJD-2440000] & $12975_{-90}^{+82}$ $^{*}$ &$19907_{-514}^{+495}$& 16548 $\pm$ 3 $^{**}$ &18389 $\pm$ 3& $5387_{-1328}^{+1173}$ $^{***}$ & $18650_{-1520}^{+1405}$ \\
        e & 0.09$_{-0.05}^{+0.18}$ &0.12 $\pm$ 0.06& 0.642 $\pm$ 0.001 &0.637 $\pm$ 0.002& 0.53 $\pm$ 0.03& 0.54 $\pm$ 0.03\\
        $\omega$ [deg] & $291.3_{-17.2}^{+40.1}$ &$288.8_{-36.4}^{+33.7}$& $-93.7_{-25.5}^{+182.8}$&330.1 $\pm$ 0.3& $2.6_{-1.4}^{+1.3}$& $3.3_{-1.3}^{+1.4}$ \\
        \textit{a} [au] & 3.53 $\pm$ 0.07 &3.54$_{-0.07}^{+0.08}$& 3.28 $\pm$ 0.04 &3.24 $\pm$ 0.03& $9.99_{-0.71}^{+0.78}$& $10.4_{-0.8}^{+0.9}$\\
        $\mathrm{M_{c}}\sin{i}$ [\mjup] & 1.55 $\pm$ 0.15 &1.6 $\pm$ 0.3 &9.89 $\pm$ 0.25 &9.7 $\pm$ 0.5& - &3.9 $\pm$ 0.4 \\ 
     \hline   
    \hline
    \end{tabular}
    \\
    \footnotesize\raggedright{{\textbf{Note.} \\
    $^{*}$ T$_{conj}$ [BJD] = $2454457_{-90}^{+82}$ \citep{endl2022jupiter}
    $^{**}$ Epoch of periastron T$_{conj}$ [BJD] = 2458388.6 $\pm$ 2.2 \citep{damasso2020precise} $^{***}$ In JD. \\
    $^{(a)}$ \cite{GDR22018}; $^{(b)}$ \cite{GEDR3}; $^{(c)}$ \cite{mermilliod1987ubv}; $^{(d)}$ \cite{gandolfi2018tess}; $^{(e)}$ \cite{schweitzer2019carmenes}; $^{(f)}$ \cite{koen2010ubv}; $^{(g)}$ \cite{stassun2019revised}; $^{(h)}$ \cite{gray2006contributions}}. All the $\Delta\mu_{\alpha}$, $\Delta\mu_{\delta}$ and $\Delta\nu_{T}$ are from \cite{kervella2022stellar}.}
    \label{tab:allvalidation}
\end{table*}

\subsection{\text{GJ 463 b}}
GJ 463 (HIP 60398) is an early-M dwarf located at a distance of 18.4 pc. \cite{endl2022jupiter} first reported the discovery of GJ 463 b identifying it as a planetary companion with a minimum mass of $\mathrm{M_{c}}\sin{i}$ $\sim$ 1.6 $\mathrm{M_{Jup}}$, through RV monitoring. 
Subsequently, \cite{sozzetti2023dynamical} conducted an analysis combining RV and PMA data to constrain \textit{i}, \textsc{$\Omega$} and $\mathrm{M_{c}}$ of the companion. Utilizing the HGCA and GDR3 catalogues, the author derived a precise mass of $\mathrm{M_{c}}$ = 3.6 $\pm$ 0.4 \mjup.
As \cite{endl2022jupiter} and \cite{sozzetti2023dynamical}, we used 70 RV data points: 53 obtained with HRS/HET \citep{Tull1998HRS} between April 2008 and June 2013, and 17 from HIRES/Keck \citep{Vogt1994HIRES} collected from January 2010 to January 2022.

\begin{figure}
\centering
\includegraphics[width=0.8\linewidth]{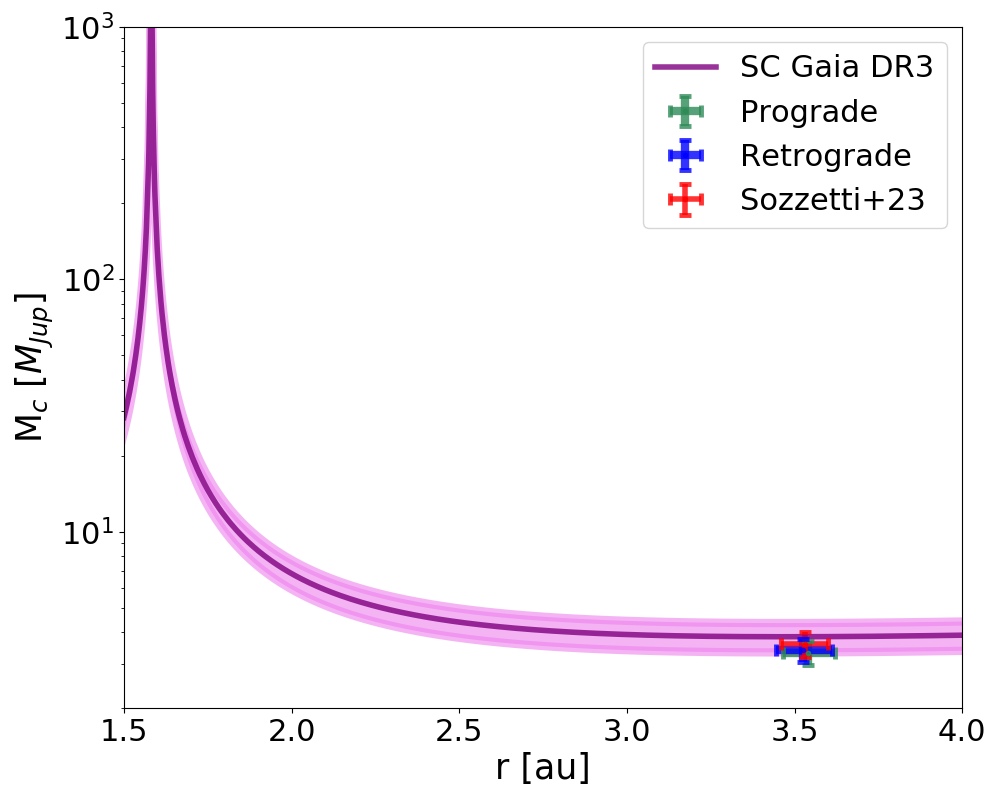}
\caption{Sensitivity Curve for GJ 463 companion. The purple line is the sensitivity curve using GDR3 with its 1$\sigma$ level of confidence. The red cross represents the value obtained by \cite{sozzetti2023dynamical} using GDR3; the green cross is the prograde value and the blue cross is the retrograde value we obtained with GDR3 in this work: the three values are compatible with the sensitivity curve.} 
\label{fig:SC_GJ463b}
\end{figure}

We applied our model using GDR3 data and we present the posterior distributions in a corner plot \citep{cornerplot} in Figure \ref{fig:CP_gj463b} and the RV offsets and the jitter terms in the Appendix (Table \ref{tab:rvoffset_gj463b}). Table \ref{tab:param_gj463b} summarizes the astrometric parameters results and the precise mass obtained by \cite{sozzetti2023dynamical} and our model: the solutions obtained in \cite{sozzetti2023dynamical} and in this work are compatible between them and strictly constrained. In Figure \ref{fig:SC_GJ463b}, we reported the mass values with the sensitivity curve. In the plot, the results from \cite{sozzetti2023dynamical} and our analysis fall within the 1$\sigma$ confidence level of the curve derived using GDR3 data, confirming the planetary nature of GJ 463 b and the accuracy of the mass value determined. The RUWE value for the GJ 463 system is 1.407, which approaches the limit where a single-star model fails to describe the data. This elevated RUWE suggests that the GDR3 time baseline detects the presence of GJ 463 b. However, since the orbital period exceeds $\delta t_{\mathrm{GDR3}}$, the resulting excess scatter is modest, implying that the astrometric data are still reasonably modelled by a single-star solution, as discussed in \cite{sozzetti2023dynamical}.

\begin{table}[]
 \caption{Derived parameters for GJ 463 b.}
 \small
    \centering
    \renewcommand\arraystretch{1.2}
    \begin{tabular}{lccc}
    \hline
    \hline
         Parameter & \cite{sozzetti2023dynamical} & This work \\
         \hline
        & Prograde solution & \\
         \hline
         $i_{p}$ [deg] & 27 $\pm$ 3 & 27.7$_{-3.9}^{+4.6}$ \\
         \textsc{$\Omega$} [deg] & 148$_{-5}^{+6}$ & 163.9 $\pm$ 7.3\\
         $\mathrm{M_{c}}$ [\mjup] & 3.4 $\pm$ 0.3 & 3.3$_{-0.3}^{+0.4}$ \\ 
         \hline
         & Retrograde solution & \\
         \hline
         $i_{r}$ [deg] & 152$_{-3}^{+2}$ & 152.3$_{-4.6}^{+3.9}$ \\
         \textsc{$\Omega$} [deg] & 80$_{-5}^{+6}$ & 67.9$_{-8.1}^{+7.9}$ \\
         $\mathrm{M_{c}}$ [\mjup] & 3.6 $\pm$ 0.4 & 3.4 $\pm$ 0.4 \\
        \hline
        \hline
             \end{tabular}
    \label{tab:param_gj463b}
\end{table}

\subsection{$\pi$ Men b}
$\pi$ Men (HD 39091) is a solar-type star hosting a BD companion $\pi$ Men b \citep{Jones2002HD39091}, a transiting super-Earth \citep[$\pi$ Men c, P $\sim$ 6.27 days; R $\sim$ 2 $\mathrm{R_{\odot}}$;][]{gandolfi2018tess} and a potential third companion with $\mathrm{P_{d}}$ $\sim$ 125 days and minimum mass of 0.042 $\pm$ 0.004 $\mathrm{M_{Jup}}$ \citep{piMend}. 
Several studies have analysed $\pi$ Men system to constrain the mass of the long-period companion using different techniques based on \HipGaia astrometry \citep{xuan2020evidence,damasso2020precise,de2020significant,venner2021true}. Our primary comparison is with Damasso et al. (\citeyear{damasso2020precise}), as the model applied in \cite{sozzetti2023dynamical} was previously used in their analysis. However, we also considered results from other works. These studies provided estimates of both the orbital inclination and the true mass of $\pi$ Men b.
\citet{damasso2020precise} conducted a joint photometric and RV analysis using TESS \citep{tess} light curves and 520 nightly binned RVs obtained from multiple instruments: UCLES/AAT \citep{ucles}, CORALIE/Euler \citep{coralie}, HARPS/ESO \citep{HARPS} and ESPRESSO/VLT \citep{espresso}. They modelled PMA using HGCA and GDR2 data, specifically to constrain the mutual inclination between the orbits of $\pi$ Men b and c. For $\pi$ Men b, they derived an orbital inclination of 45.8$_{-1.1}^{+1.4}$° and a true mass of 14.1$_{-0.4}^{+0.5}$ \mjup. 

In our analysis, we used 235 RV measurements: 42 from UCLES/AAT (taken from November 1998 to October 2005), 49 from CORALIE (from October 1999 to February 2020), and 144 by HARPS published in \cite{gandolfi2018tess} (from December 2003 to March 2016). The ESPRESSO data were excluded as they are not publicly available.
To account for the instrumental upgrades, we treated the HARPS data as two separate datasets (HARPS ``pre'' and HARPS ``post'') because on June 2015 the HARPS fiber bundle was upgraded possibly introducing an RV offset \citep{2015Msngr.162....9L}. CORALIE data were divided into three datasets, due to two upgrades that occurred in 2007 and 2014, resulting in a small RV offset between the datasets \citep{coralie07}. We used GDR3 data to improve parameter constraints. For the RV analysis, we did not considered the two inner companions, as their signals would be absorbed in the jitter terms ($\mathrm{K_{c}}$ = 1.5 $\pm$ 0.2 $\mathrm{ms^{-1}}$, $\mathrm{K_{d}}$ = 1.68 $\pm$ 0.17 $\mathrm{ms^{-1}}$), as also highlighted in \cite{xuan2020evidence}.

\begin{figure}
\centering
\includegraphics[width=.8\linewidth]{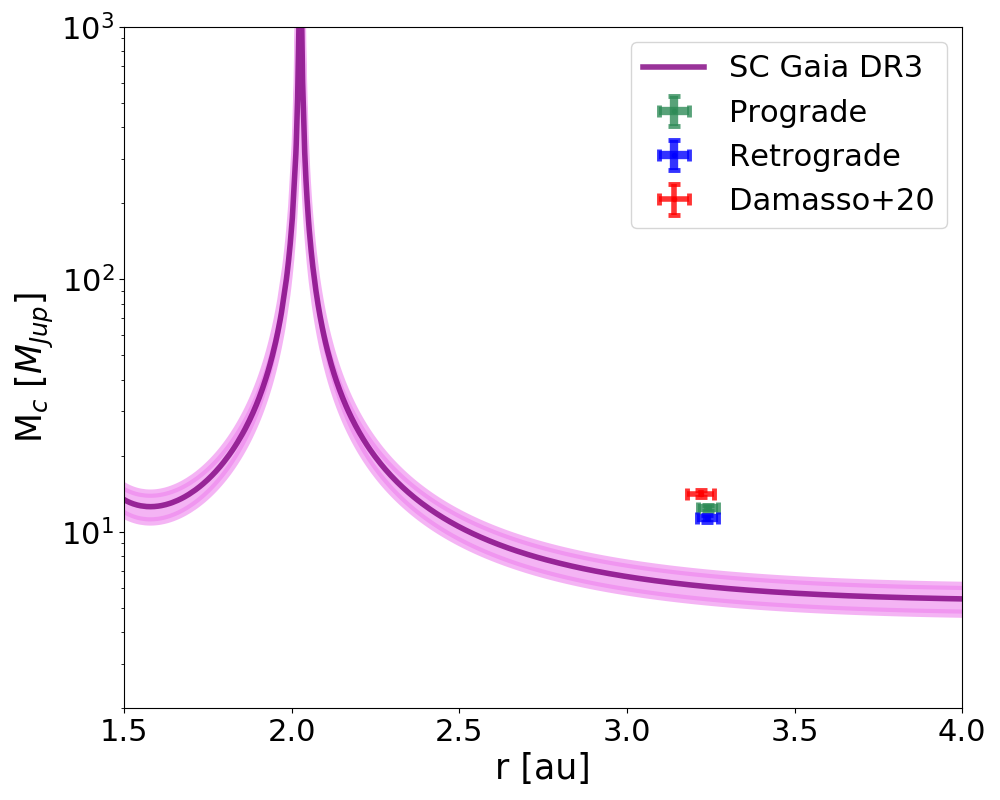}
\caption{Sensitivity Curve for $\pi$ Men companion. The red cross represents the value obtained by \cite{damasso2020precise} using GDR2; the green and the blue crosses are the results we obtained with GDR3 in this work for the prograde and the retrograde solutions, respectively: the values stay over the sensitivity curve.} 
\label{fig:SC_piMenb}
\end{figure}

We obtained true masses of 12.4 $\pm$ 0.3 $\mathrm{M_{Jup}}$ and 11.3$_{-0.4}^{+0.3}$ \mjup, for the prograde and retrograde solutions, respectively. These results lie at the boundary between the planetary and BD regime. The RV offsets and the jitter terms are reported in the Appendix (Tab. \ref{tab:rvoffset_piMenb}) with the posterior distributions in Figure \ref{fig:CP_piMenb}. The astrometric solutions and the mass values are reported in Table \ref{tab:param_piMenb}.
The sensitivity curve (Fig. \ref{fig:SC_piMenb}) indicates that both the masses determined by \cite{damasso2020precise} and in this work reside above the curve. This is attributable to $\pi$ Men b's periastron passage around 2013 and 2018, which lie outside the observational windows of GDR3 (2014.6-2017.4). Consequently, \Gaia observed the apastron passage, where the tangential velocity anomaly is minimal, limiting astrometric sensitivity to the companion's mass.
Although our inclinations are less precise compared to Damasso et al. (\citeyear{damasso2020precise}), we achieved comparable mass precision using fewer RV data points (235 vs 520). Combining RV datasets (UCLES/AAT and HARPS points) with astrometric data from GDR2, \citet{xuan2020evidence}, \citet{de2020significant} and \citet{venner2021true} (the latter utilizing re-reduced HARPS RVs from \citealt{Trifonov2020}) also determined the orbital inclination and precise mass of $\pi$ Men b. Our spectroscopic results are consistent with their results, despite the RV points are slightly different. 
As reported in Table \ref{tab:param_piMenb}, the different $\Omega$ results are due to the fact that with the PMA method we use only two values to derive the orbit, so it is not possible to strictly constrain it and its direction. Depending on how the orbit is simulate, the direction obtained could be different, such $\Omega$ value obtained by \citet{xuan2020evidence}, \citet{damasso2020precise} and \citet{venner2021true} and the $\Omega$ value obtained by \citet{de2020significant} and in this work. 
\begin{table*}[]
\small
 \caption{Derived parameters for $\pi$ Men b.}
    \centering
    \renewcommand\arraystretch{1.2}
    \begin{tabular}{lccccc}
    \hline
    \hline
     Parameter & \cite{xuan2020evidence} & \cite{damasso2020precise} & \cite{de2020significant} & \cite{venner2021true} & This work \\
     \hline
     && Prograde solution &&& \\
     \hline
     $i_{p}$ [deg] & $51.2_{-9.8}^{+14.1}$ & $45.8_{-1.1}^{+1.4}$ & $49.9_{-4.5}^{+5.3}$ & $50.8_{-8.8}^{+12.5}$ & $51.1_{-2.8}^{+3.0}$ \\
     \textsc{$\Omega$} [deg] & $105.80_{-14.30}^{+15.08}$ & 108.8$_{-0.7}^{+0.6}$ & $270.3_{-8.0}^{+8.1}$ & $106.8_{-15.0}^{+14.9}$ & 282.5 $_{-4.2}^{+4.3}$ \\
     $\mathrm{M_{c}}$ [\mjup] & $12.9_{-1.9}^{+2.3}$ & 14.1$_{-0.4}^{+0.5}$ & $13.01_{-0.95}^{+1.03}$ & $12.9_{-1.7}^{+2.1}$ & 12.4 $\pm$ 0.3 \\ 
     \hline
     && Retrograde solution &&& \\
     \hline
     $i_{r}$ [deg] & - & - & - & - & $120.9_{-4.8}^{+3.9}$ \\
     \textsc{$\Omega$} [deg] & - & - & - & - & $41.3_{-8.8}^{+7.0}$ \\
     $\mathrm{M_{c}}$ [\mjup] & - & - & - & - & 11.3$_{-0.4}^{+0.3}$ \\
    \hline
    \hline
    \end{tabular}
    \label{tab:param_piMenb}
\end{table*}

\subsection{HD 222237 b}
\begin{figure}
\centering
\includegraphics[width=.8\linewidth]{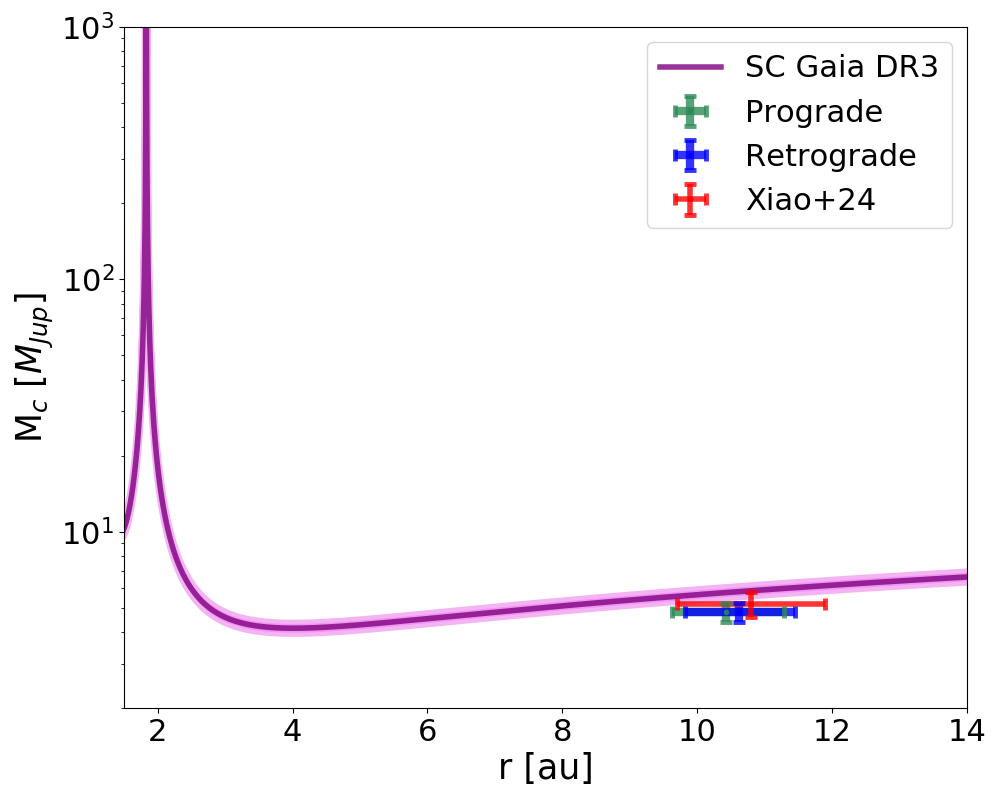}
\caption{Sensitivity Curve for HD 222237 companion. The red cross represents the value obtained by \cite{xiao2024hd} using GDR3; the green cross is the prograde value and the blue cross is the retrograde value we obtained with GDR3 in this work: the three values are compatible with the sensitivity curve.} 
\label{fig:SC_HD222237b}
\end{figure}

The K3 dwarf HD 222237 (HIP 116745) hosts a super-Jupiter first reported by \cite{xiao2024hd}. In their study, the authors compared the performance of the RV+PMA model implemented in the \texttt{orvara} tool \citep{orvara} with their new technique. \texttt{orvara} is designed to fit full orbital parameters using any combination of RVs, relative and absolute astrometry, using HGCA, which corresponds to a single \Gaia data release. Instead, \cite{xiao2024hd} combined \Hipparcos epoch data with GDR2 and GDR3, in order to break the degeneracy between prograde and retrograde orbital inclinations. Their analysis utilized 63 RV data points from the PSF/Magellan II \citep{PSF}, 35 from UCLES/AAT, and 54 from HARPS, distinguished in HARPS ``pre'' and HARPS ``post''. This approach enable them to determine the true orbital inclination $49.9_{-2.8}^{+3.4}$°, defining the orbit as prograde, with a precise mass of 5.19 $\pm$ 0.58 \mjup.
We conducted the analysis of this system using the same RV data. As reported in Table \ref{tab:param_HD222237b}, we evaluate \textit{i} as $55.3_{-3.0}^{+4.4}$° and $125.0_{-4.3}^{+2.8}$°, for two solutions, and a mass of $\sim$4.8 \mjup. Figure \ref{fig:SC_HD222237b} shows that the Xiao's mass and our results are compatible with the sensitivity curve, confirming the effectiveness of our procedure. We reported the posterior distributions in Figure \ref{fig:CP_HD222237b}, along with the RV offsets and the jitter terms (Tab \ref{tab:rvoffset_HD222237b}), in the Appendix.

\begin{table*}[]
\small
 \caption{Derived parameters for HD 222237 b.}
    \centering
    \renewcommand\arraystretch{1.2}
    \begin{tabular}{lccc}
    \hline
    \hline
         Parameter & \cite{xiao2024hd} (RV+HG3) & \cite{xiao2024hd} (RV+HG23) & This work \\
         \hline
         &Prograde solution&&\\
         \hline
         $i_{p}$ [deg] & 57.8$_{-3.8}^{+4.9}$ & $49.9_{-2.8}^{+3.4}$ & 55.3$_{-3.0}^{+4.4}$ \\
         \textsc{$\Omega$} [deg] & $76.0_{-6.1}^{+7.4}$ & $69.8_{-5.7}^{+6.7}$ & $69.9_{-5.7}^{+7.4}$ \\
         $\mathrm{M_{c}}$ [\mjup] & 4.56$_{-0.49}^{+0.51}$ & 5.19 $\pm$ 0.58 & 4.78 $\pm$ 0.42 \\ 
         \hline
         &Retrograde solution&&\\
         \hline
         $i_{r}$ [deg] & 122.9$_{-4.5}^{+3.3}$ & - & $125.0_{-4.3}^{+2.8}$ \\
         \textsc{$\Omega$} [deg] & 135.3$_{-6.9}^{+5.6}$ & - & $139.7_{-7.2}^{+5.3}$\\
         $\mathrm{M_{c}}$ [\mjup] & - & - & $4.81_{-0.43}^{+0.41}$ \\
        \hline
        \hline
             \end{tabular}
    \label{tab:param_HD222237b}
\end{table*}

The results obtained by applying our model to these three objects are consistent with those reported in the literature, confirming the accuracy and validation of our code.

\section{Results}
\label{sec:Results}
We consider the nine objects analysed by \cite{kiefer2021determining} in order to evaluate their inclinations and masses. Since the work was made using \Gaia DR1 to simulate the astrometric noise, we decided to study these objects with new \Gaia data releases applying our method. Considering that \Gaia and, in particular astrometry, is more efficient for objects with P $>$ 1 year, HD 114762 system is excluded from our sample (HD 114762 b, P = 83.913 $\pm$ 0.003 days). The stellar properties of the eight systems are reported in Table \ref{tab:allsystems} and the results of the systems analysed are in Table \ref{tab:allresults}.

\begin{table*}[h!]
   \centering
   \tiny
   \caption{Stellar properties of the eight planetary systems}
    \renewcommand\arraystretch{1.2}
    \begin{tabular}{l|cccc}
    \hline
    \hline
       Parameter & HD 5388 $^{(a,b,c)}$ & HD 6718 $^{(a,b,c)}$ & HD 16760 $^{(a,b,d,e)}$ & 30 Ari B $^{(a,f)}$ \\
    \hline
        RA [J2000] & 00:55:11.89 & 01:07:48.66 & 02:42:21.31 & 02:36:57.75\\
        Dec [J2000] & -47:24:21.48 & -08:14:01.33 & +38:37:07.23 & +24:38:53.00\\
        Parallax $\pi$ [mas] & 8.84 $\pm$ 0.02 & 19.45 $\pm$ 0.02 & 17.57 $\pm$ 0.59 & 22.51 $\pm$ 0.03\\
        Distance \textit{d} [pc] & 53.07 $\pm$ 0.06 & 51.40 $\pm$ 0.06 & 56.8 $\pm$ 1.9 & 44.42 $\pm$ 0.06\\
        $m_{V}$ & 6.72 $\pm$ 0.01 & 8.45 $\pm$ 0.01 & 8.76 $\pm$ 0.01 & 7.3\\
        Radius $\mathrm{R_{\star}}$ [$R_{\odot}$] & 1.91 & 1.02 $\pm$ 0.03 & 0.835 $\pm$ 0.005 & 1.13 $\pm$ 0.03\\
        Mass $\mathrm{M_{\star}}$ [$M_{\odot}$] & 1.21 & 0.96 & 0.93 $\pm$ 0.01 & 1.16 $\pm$ 0.04\\
        Effective temperature $\mathrm{T_{eff}}$ [K] & 6297 $\pm$ 32 & 5746 $\pm$ 19 & 5518 $\pm$ 11 & 6152\\
        Spectral type & F6V & G5V & G5V & F6V\\
        \hline
        $\Delta\mu_{\alpha}$ [mas yr$^{-1}$] (\textsc{Hipparcos}) & 0.193 $\pm$ 0.550 & -0.421 $\pm$ 0.820 & 2.309 $\pm$ 2.241 & 3.701 $\pm$ 0.750\\
        $\Delta\mu_{\delta}$ [mas yr$^{-1}$] (\textsc{Hipparcos}) & 0.072 $\pm$ 0.490 & -0.266 $\pm$ 0.510 & -0.683 $\pm$ 2.390 & -0.444 $\pm$ 0.540\\
        $\Delta\mu{_\alpha}$ [mas yr$^{-1}$] (\textit{Gaia} DR3) & -0.069 $\pm$ 0.021 & -0.073 $\pm$ 0.037 & -3.051 $\pm$ 0.693 & -6.071 $\pm$ 0.037\\
        $\Delta\mu_{\delta}$ [mas yr$^{-1}$] (\textit{Gaia} DR3) & 0.088 $\pm$ 0.024 & 0.068 $\pm$ 0.031 & 0.213 $\pm$ 0.699 & 1.825 $\pm$ 0.032\\
        $\Delta\nu_{T}$ [m s$^{-1}$] (\textit{Gaia} DR3) & 28.02 $\pm$ 8.11 & 24.26 $\pm$ 11.83 & 823.45 $\pm$ 265.03 & 1332.82 $\pm$ 10.29\\
        RUWE & 1.130 & 0.957 & 26.546 & 1.398\\
     \hline   
    \hline
       Parameter & HD 141937 $^{(a,b,g,h)}$ & HD 148427 $^{(a,b,g)}$ & HD 96127 $^{(a,b,g)}$ & HIP 65891 $^{(a,c,i)}$\\
    \hline
        RA [J2000] & 15:52:17.65 & 16:28:28.15 & 11:05:45.95 & 13:30:25.27\\
        Dec [J2000] & -18:26:09.50 & -13:23:58.69 & +44:18:05.61 & -58:39:51.68\\
        Parallax $\pi$ [mas] & 30.56 $\pm$ 0.09 & 14.21 $\pm$ 0.02 & 1.87 $\pm$ 0.02 & 6.70 $\pm$ 0.02\\
        Distance \textit{d} [pc] & 32.72 $\pm$ 0.10 & 70.4 $\pm$ 0.1 & 535.3 $\pm$ 6.2 & 149.3 $\pm$ 0.4\\
        $m_{V}$ & 7.25 $\pm$ 0.01 & 6.89 $\pm$ 0.01 & 7.41 $\pm$ 0.01 & 6.77\\
        Radius $\mathrm{R_{\star}}$ [$R_{\odot}$] & 1.03 & 3.86 $\pm$ 0.12 & 35 $\pm$ 17  & 8.93 $\pm$ 1.02\\
        Mass $\mathrm{M_{\star}}$ [$M_{\odot}$] & 1.09 & 1.64 $\pm$ 0.27 & 0.91 $\pm$ 0.25 & 2.50 $\pm$ 0.21\\
        Effective temperature $T_{\text{eff}}$ [K] & 5870 $\pm$ 9 & 5025 $\pm$ 4 & 3943 $\pm$ 23 & 5000 $\pm$ 100\\
        Spectral type & G1V & K0III/IV & K2III & K0III\\
        \hline
        $\Delta\mu_{\alpha}$ [mas yr$^{-1}$] (\textsc{Hipparcos}) & -0.812 $\pm$ 0.770 & -0.328 $\pm$ 0.760 & -0.439 $\pm$ 0.470 & 0.650 $\pm$ 0.530 \\
        $\Delta\mu_{\delta}$ [mas yr$^{-1}$] (\textsc{Hipparcos}) & 0.391 $\pm$ 0.560 & -0.251 $\pm$ 0.510 & -0.079 $\pm$ 0.440 & 0.503 $\pm$ 0.400\\
        $\Delta\mu{_\alpha}$ [mas yr$^{-1}$] (\textit{Gaia} DR3) & 0.243 $\pm$ 0.104 & 0.019 $\pm$ 0.034 & 0.007 $\pm$ 0.022 & 0.045 $\pm$ 0.057$^{*}$\\
        $\Delta\mu_{\delta}$ [mas yr$^{-1}$] (\textit{Gaia} DR3) & -0.219 $\pm$ 0.069 & -0.008 $\pm$ 0.020 & -0.062 $\pm$ 0.033 & 0.064 $\pm$ 0.064$^{*}$\\
        $\Delta\nu_{T}$ [m s$^{-1}$] (\textit{Gaia} DR3) & 50.67 $\pm$ 19.37 & 6.94 $\pm$ 13.01 & 154.34 $\pm$ 98.20 & 56.14 $\pm$ 61.01$^{*}$\\
        RUWE & 4.972 & 1.055 & 0.997 & 0.904\\
        \hline
        \hline   
    \end{tabular}
    \\
    \footnotesize\raggedright{{\textbf{Note.} \\ 
    $^{(a)}$ \cite{GEDR3}; $^{(b)}$ \cite{hog2001tycho}; $^{(c)}$ \cite{houk1982university}; $^{(d)}$ \cite{bonfanti2015revising}; $^{(e)}$ \cite{hipparcos}; $^{(f)}$ \cite{nordstrom1997radial}, $^{(g)}$ \cite{stassun2017accurate}; $^{(h)}$ \cite{gray2006contributions}; $^{(i)}$ \cite{loden1967photometric}. All the $\Delta\mu_{\alpha}$, $\Delta\mu_{\delta}$ and $\Delta\nu_{T}$ are from \cite{kervella2022stellar}.} $^{*}$ $\Delta\mu_{\alpha}$, $\Delta\mu_{\delta}$ and $\Delta\nu_{T}$ are from \cite{kervella2019stellar}, for GDR2}
    \label{tab:allsystems}
    \end{table*}

\begin{table*}[h!]
\fontsize{7.5}{12}\selectfont
   \centering
   \caption{Properties of the eight planetary systems analysed.}
    \renewcommand\arraystretch{1}
    \begin{tabular}{l|cccccccc}
    \hline
    \hline
       Parameter & HD 5388 b & & HD 6718 b & & HD 16760 b & & 30 Ari B b&\\
        \hline
        &Santos (\citeyear{santos2010harps})&This Work&Naef (\citeyear{NaefHD6718})&This Work&Sato (\citeyear{sato2009substellar})&This Work& Kane (\citeyear{2015ApJ...815...32K})&This Work\\
        \textsc{K} [ms$^{-1}$]&41.7 $\pm$ 1.6&$42.7_{-1.3}^{+1.4}$&24.1 $\pm$ 1.5&$28.4_{-0.9}^{+1.0}$&407.7 $\pm$ 0.8& 407.0 $\pm$ 0.9&177 $\pm$ 26 & $281_{-25}^{+27}$\\
        \textsc{P} [days]&777 $\pm$ 4& 772.5 $\pm$ 1.0&2496 $\pm$ 176&2476 $\pm$ 19&466.5 $\pm$ 0.4&466.08 $\pm$ 0.08&345.4 $\pm$ 3.8&339.7$_{-0.2}^{+0.3}$\\
        $\mathrm{T_{p}}$ [BJD-2450000]&4570 $\pm$ 9&$8432_{-23}^{+24}$&4357 $\pm$ 251&$8193_{-170}^{+208}$&3337.0 $\pm$ 2.4 $^{**}$&7995 $\pm$ 4&3222.1 $\pm$ 42.4& 8000.5$_{-16.0}^{+13.6}$ \\
        e&0.40 $\pm$ 0.02&0.41 $\pm$ 0.02&$0.10_{-0.04}^{+0.11}$&0.06 $\pm$ 0.03&0.084 $\pm$ 0.003&0.079 $\pm$ 0.002&0.18 $\pm$ 0.11&0.45 $\pm$ 0.04\\
        \textsc{$\omega$} [deg]&324 $\pm$ 4&$325.9_{-2.7}^{+2.8}$&$286_{-35}^{+64}$&$129_{-27}^{+34}$&242.9 $\pm$ 1.9&$240.4_{-1.7}^{+1.6}$&337 $\pm$ 57&354$_{-9}^{+8}$\\
        \textit{a}$^{*}$ [au]&1.76 $\pm$ 0.09&1.757 $\pm$ 0.001&$3.56_{-0.15}^{+0.24}$&3.53 $\pm$ 0.05&1.08 $\pm$ 0.02&1.156 $\pm$ 0.004&1.01 $\pm$ 0.01&1.05 $\pm$ 0.01\\
        $\mathrm{M_{c}}\sin{i}$$^{*}$ [\mjup]&1.96 &$2.0_{-0.3}^{+0.5}$&1.6 $\pm$ 0.1&$1.8_{-0.3}^{+0.5}$&13.1 $\pm$ 0.6&$14.8_{-7.6}^{+7.8}$&6.6 $\pm$ 0.9&9.5 $\pm$ 1.3\\ 
       \hline
            &&&&Prograde solution&&&&\\
       \hline
           &Kiefer (\citeyear{kiefer2021determining})&This Work&Kiefer (\citeyear{kiefer2021determining})&This Work& Kiefer (\citeyear{kiefer2021determining})&This Work&Kiefer (\citeyear{kiefer2021determining})&This Work\\
       $i_{p}$ [deg]&1.4 $\pm$ 0.2&$38.7_{-6.8}^{+10.2}$&$1.488_{-0.310}^{+0.410}$&$51.4_{-11.5}^{+16.4}$&$3.164_{-0.762}^{+0.810}$&$43.3_{-28.8}^{+32.2}$&$4.181_{-0.931}^{+1.031}$&2.9 $\pm$ 0.3\\
       \textsc{$\Omega$} [deg]&-&$266.6_{-12.8}^{+13.4}$&-&$234.7_{-24.8}^{+28.9}$&-&$84.6_{-36.5}^{+53.4}$&-& 39.0$_{-6.3}^{+6.4}$\\
       $\mathrm{M_{c}}$$^{*}$ [\mjup]&$87_{-11}^{+14}$&$3.2_{-0.2}^{+0.3}$&$62.79_{-13.80}^{+16.98}$&$2.4_{-0.2}^{+0.3}$&$219.9_{-69.4}^{+120.7}$&$21.5_{-4.5}^{+5.0}$&$147.4_{-29.5}^{+41.3}$&188.1$_{-18.5}^{+19.7}$\\ 
       \hline
           &&&&Retrograde solution&&&&\\
       \hline
            &Sahlmann (\citeyear{sahlmann2011hd})&This Work&&This Work&&This Work&&This Work\\
       $i_{r}$ [deg]&$178.3_{-0.7}^{+0.4}$&$142.4_{-8.5}^{+6.2}$&-&$129.0_{-16.2}^{+11.2}$&-&$132.6_{-29.6}^{+30.3}$&-&177.1 $\pm$ 0.3\\
       \textsc{$\Omega$} [deg]&$298.0_{-26.5}^{+16.4}$&$17.8_{-10.5}^{+11.5}$&-&$42.6_{-22.4}^{+31.7}$&-&$107.0_{-42.9}^{+47.7}$&-&159.4$_{-6.3}^{+6.2}$\\
       $\mathrm{M_{c}}$$^{*}$ [\mjup]&69.2 $\pm$ 19.9&3.3 $\pm$ 0.2&-&$2.4_{-0.3}^{+0.2}$&-&$20.0_{-4.0}^{+4.1}$&-&184.0$_{-17.9}^{+18.6}$\\
       \hline
       \hline
       Parameter&HD 141937 b&&HD 148427 b&&HD 96127 b&&HIP 65891 b&\\
       \hline
        &Udry (\citeyear{udry2002coralie})&This Work&Fischer (\citeyear{Fischer2009HD148427b})&This Work&Gettel (\citeyear{gettel2012hd96127b})&This Work&Jones (\citeyear{jones2015hip65891})&This Work\\
       \textsc{K} [ms$^{-1}$]&234.5 $\pm$ 6.4& 278.4 $\pm$ 2.2&27.7 $\pm$ 2.0&$25.1_{-2.5}^{+2.4}$&104.8 $\pm$ 10.6&$100.7_{-14.4}^{+19.5}$&64.9 $\pm$ 2.4&$65.7_{-2.4}^{+2.5}$\\
       \textsc{P} [days]&653.22 $\pm$ 1.21&662.37 $\pm$ 0.09&331.5 $\pm$ 3.0&$331.9_{-2.2}^{+1.9}$&647.3 $\pm$ 16.8&$638.9_{-16.7}^{+17.0}$&1084.5 $\pm$ 23.2&$1089.6_{-22.0}^{+22.8}$\\
       $\mathrm{T_{p}}$ [BJD-2450000]&1847.38 $\pm$ 1.97 $^{**}$&8486 $\pm$ 1 &3991 $\pm$ 15 $^{**}$&$7969_{-85}^{+83}$&3969.4 $\pm$ 31.0&$8441_{-24}^{+27}$&6014.8 $\pm$ 49.3 $^{**}$&8220.9$_{-73.2}^{+87.8}$\\
       e&0.41 $\pm$ 0.01&0.460 $\pm$ 0.004&0.16 $\pm$ 0.08&$0.12_{-0.08}^{+0.09}$&0.3 $\pm$ 0.1&0.3 $\pm$ 0.1&0.13 $\pm$ 0.05&0.09 $\pm$ 0.05\\
       \textsc{$\omega$} [deg]&187.72 $\pm$ 0.80&192.7 $\pm$ 0.6&277 $\pm$ 68&$267.2_{-68.4}^{+40.3}$&162.0 $\pm$ 18.2&$163.1_{-15.6}^{+15.9}$&355.5 $\pm$ 15.5&$363.8_{-18.5}^{+24.6}$\\
       \textit{a}$^{*}$ [au]&1.52&1.54 $\pm$ 0.02&0.93 $\pm$ 0.01&1.11 $\pm$ 0.06&1.42 $\pm$ 0.13&1.41 $\pm$ 0.13&2.81 $\pm$ 0.09&2.82 $\pm$ 0.09\\
       $\mathrm{M_{c}}\sin{i}$$^{*}$ [\mjup]&9.7&11.3 $\pm$ 0.5&0.96 $\pm$ 0.10&$1.2_{-0.3}^{+0.2}$&4.0 &$3.8_{-1.2}^{+1.3}$&6.00 $\pm$ 0.49& 6.1 $\pm$ 1.3\\ 
       \hline
           &Kiefer (\citeyear{kiefer2021determining})&This Work&Kiefer (\citeyear{kiefer2021determining})&This Work&Kiefer (\citeyear{kiefer2021determining})&This Work&Kiefer (\citeyear{kiefer2021determining})&This Work\\
        \textit{i} [deg]&$20.52_{-4.16}^{+12.47}$&$90.00_{-6.76}^{+6.75}$&$0.51_{-0.11}^{+0.16}$&90.2$_{-30.7}^{+30.5}$&$1.364_{-0.763}^{+38.527}$&89.9 $\pm$ 41.4&$1.184_{-0.207}^{+0.256}$&89.9$_{-39.9}^{+39.8}$\\
        \textsc{$\Omega$} [deg]&-&$291.5$ $_{-7.1}^{+7.0}$&-&$132.7_{-60.4}^{+114.3}$&-&$350.9_{-70.7}^{+71.6}$&-&$204.6_{-111.3}^{+89.4}$\\
        $\mathrm{M_{c}}$$^{*}$ [\mjup]&$27.42_{-9.86}^{+6.78}$& 11.3 $\pm$ 0.5 &$136.5_{-33.7}^{+37.2}$&$1.2_{-0.3}^{+0.2}$&$190.2_{-184.0}^{+284.1}$&$3.8_{-1.2}^{+1.3}$&$312.3_{-57.4}^{+74.2}$&6.1 $\pm$ 1.3\\ 
        \hline
        \hline 
    \end{tabular}
    \footnotesize\raggedright{{\textbf{Note.} \\ HD 148427 spectroscopic parameters were derived in \cite{Fischer2009HD148427b} using distance = 59.3 pc. \\
    The parameters with $^{*}$ were derived. $^{**}$ Julian Date.}}\\
    \label{tab:allresults}
\end{table*}

\subsection{HD 5388 b}
\begin{figure}
\centering
\includegraphics[width=1\linewidth]{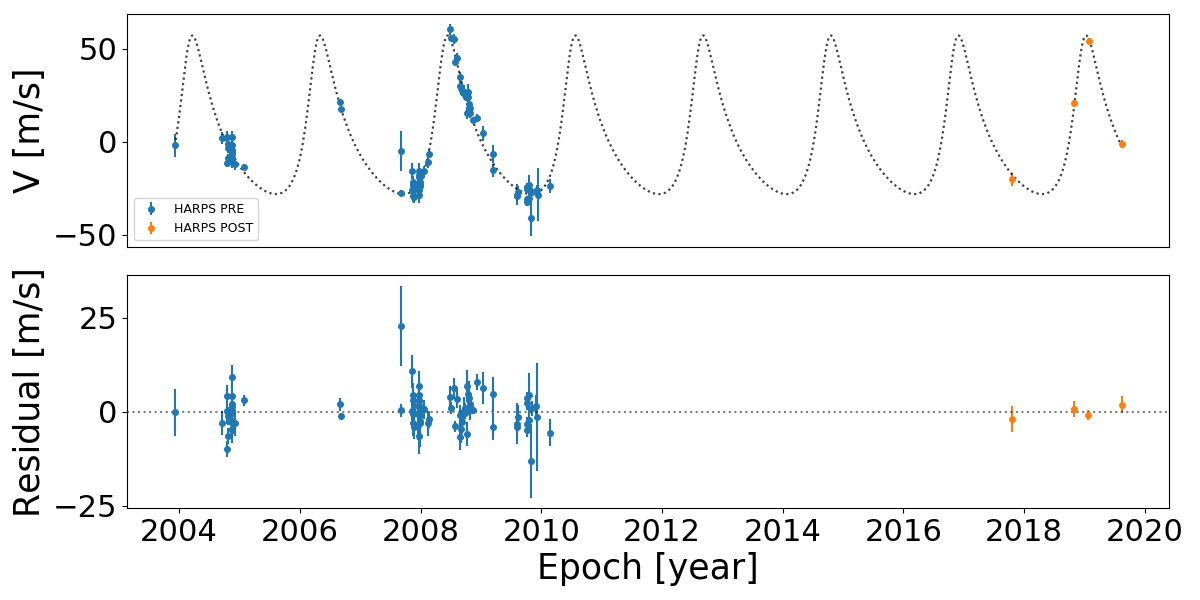}
\caption{\textit{Top:} RV data and Keplerian function of HD 5388 system. \textit{Bottom:} Residuals of the fitted orbit.}
\label{fig:RV_HD5388b}
\end{figure}

\begin{figure}
\centering
\includegraphics[width=.8\linewidth]{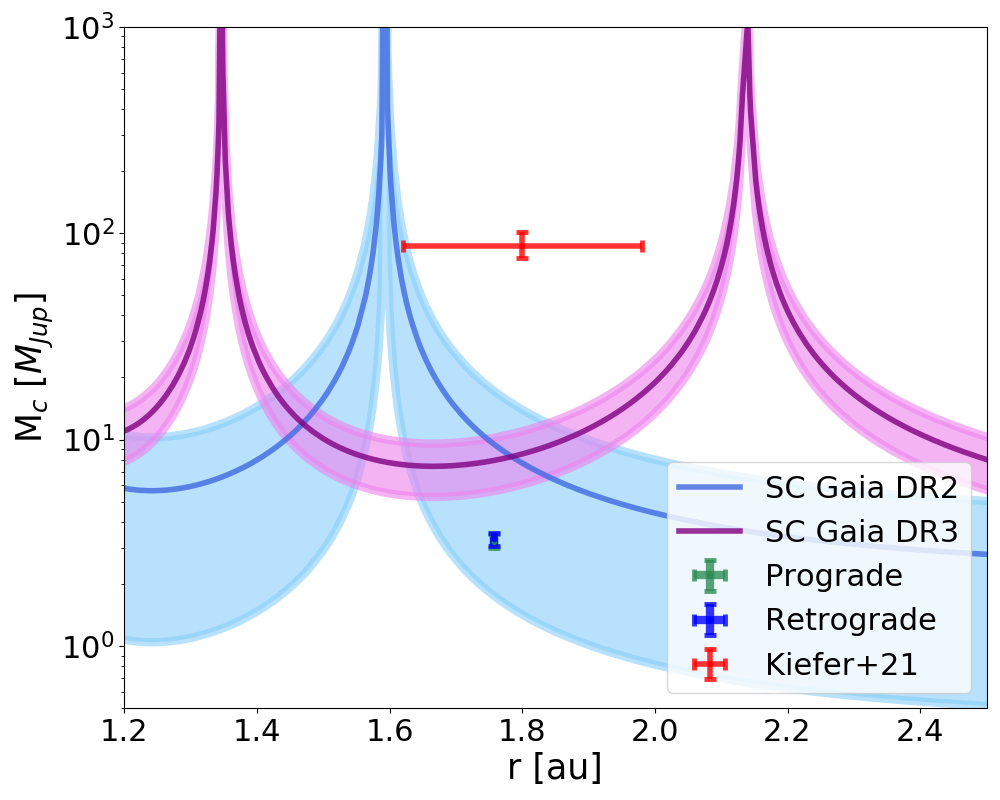}
\caption{Sensitivity Curve for HD 5388 companion. The blue line is the sensitivity curve derived using GDR2 with its 1$\sigma$ level of confidence. The red cross represents the value obtained by \cite{kiefer2021determining} using \texttt{GASTON}. The green cross represents the prograde value, while the blue cross corresponds to the retrograde value we obtained using GDR3 in this work: our results are compatible with the sensitivity curve.} 
\label{fig:SC_HD5388b}
\end{figure}

HD 5388 is a F6-type star hosting a substellar object discovered by \cite{santos2010harps} using 68 RV measurements collected by HARPS between 2003 and 2009. Subsequently, \cite{sahlmann2011hd} studied the system using intermediate astrometric data (IAD) from the \Hipparcos reduction (described in \cite{sahlmann2011search}) and found an orbital inclination of 178° and a true mass of 69.2 $\pm$ 19.9 \mjup. Later, \cite{kiefer2021determining} obtained a mass of 87.02 \mjup, constraining it within the BD to M-dwarf domain. In our study, we used 89 RV points obtained from the ESO archive for the HARPS instrument, covering the period from 2003 to 2019, including data by \cite{santos2010harps}. We compared the \cite{santos2010harps} dataset with the standard reduction provided by the ESO archive data to verify their consistency. 
Having identified only an offset between the two datasets, we decided to utilize the ESO archive data for our analysis. 
For RV data, we distinguished between HARPS ``pre'' and HARPS ``post'' datasets. We performed an RV modelling by using a Keplerian function with the set-up described in Section \ref{sec:methods}. Figure \ref{fig:RV_HD5388b} shows the RV fit while the posterior distributions and the RV offset and jitter terms are reported in the Appendix (Fig. \ref{fig:CP_HD5388b}, Tab. \ref{tab:rvoffset_HD5388b}, respectively). We derived a mass of 3.2$_{-0.2}^{+0.3}$ \mjup, for the prograde solution and 3.3 $\pm$ 0.2 $\mathrm{M_{Jup}}$ for the retrograde solution, classifying the companion as a planet.
The RUWE value of 1.130, reported in Table \ref{tab:allsystems}, indicates that the GDR3 baseline likely detected HD 5388 b, although the value remains slightly below the recommended threshold: the associated excess noise would be expected to be substantially higher, in the case of a mass of $\sim$70 \mjup.
Finally, the mass falls below the sensitivity curve representing the astrometric noise evaluated with GDR3 data (in purple), as shown in Figure \ref{fig:SC_HD5388b}. Since the mass value is below the GDR3 curve, but within 1$\sigma$ of the GDR2 area, we also conducted the analysis using GDR2 to assess any differences. We obtained consistent results with those derived from GDR3. 
Based on the comparison with the sensitivity curve, the presence of an additional sub-Jovian companion cannot be ruled out. This is further supported by a residual signal observed in the RV fit, which we analysed using the Generalised Lomb-Scargle periodogram \citep[GLS;][]{zechmeister2009generalised}, which fits the residuals with a sinusoidal function. The analysis revealed a signal with \textsc{K} = 13.2 $\pm$ 0.84 ms$^{-1}$ and \textsc{P} = 392.5 $\pm$ 1.7 days, corresponding to a minimum mass of less than 0.54 \mjup. The available data are insufficient to warrant further analysis.

\subsection{HD 6718 b}
\begin{figure}
\centering
\includegraphics[width=1\linewidth]{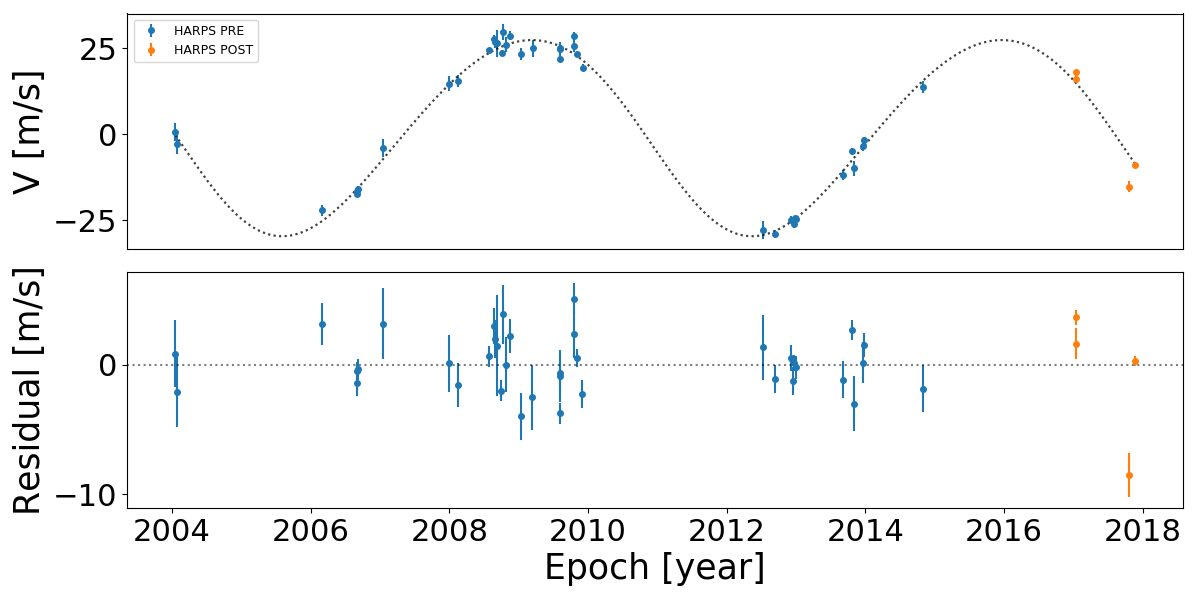}
\caption{\textit{Top:} RV data and Keplerian function of HD 6718 system. \textit{Bottom:} Residuals of the fitted orbit.}
\label{fig:RVs_HD6718b}
\end{figure}

\begin{figure}
\centering
\includegraphics[width=.8\linewidth]{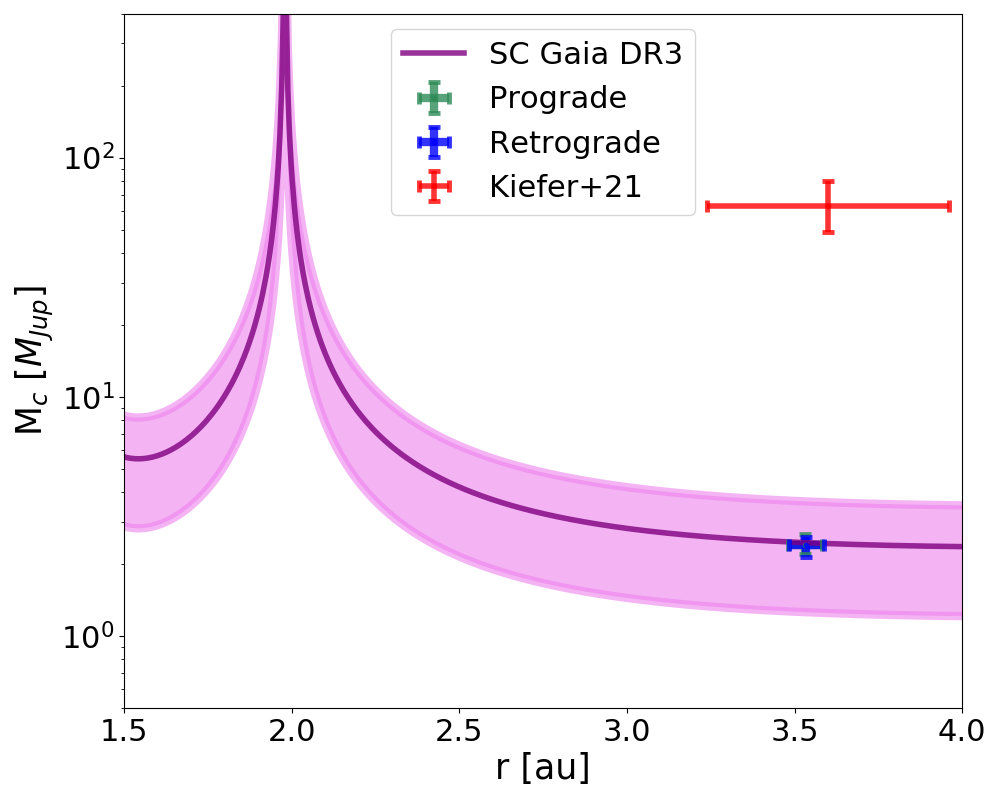}
\caption{Sensitivity Curve for HD 6718 companion. The red cross represents the value obtained by \cite{kiefer2021determining} using \texttt{GASTON}; the green cross is the prograde value and the blue cross is the retrograde value we obtained with GDR3 in this work: our results are compatible with the sensitivity curve.}
\label{fig:SC_HD6718b}
\end{figure}

HD 6718 is a G5 star hosting a companion first introduced by \cite{NaefHD6718}. The authors analysed 22 HARPS RV measurements collected between 2007 and 2009. Thereafter, \cite{kiefer2021determining}, utilizing the \texttt{GASTON} tool, derived a mass of $62.79_{-13.80}^{+16.98}$ \mjup, classifying it within the brown dwarf/low-star mass regime. In this study, we analysed 42 RV measurements, adding 20 HARPS observations from \cite{Trifonov2020}, covering a total time span from 2007 to 2017. The data was divided into HARPS ``pre'' and HARPS ``post'' datasets. The data from \cite{NaefHD6718} are included in the HARPS ``pre'' dataset. In order to obtain a homogeneous dataset, we used the data reported in \cite{Trifonov2020}, which were analysed using the \texttt{SERVAL} pipeline \citep{serval}.
The RV data and model are displayed in Figure \ref{fig:RVs_HD6718b}. The posterior distributions (Figure \ref{fig:CP_HD6718b}) and the RV offsets and jitter terms (Table \ref{tab:rvoffset_HD6718b}) are reported in the appendix. We derived a mass of $\sim$2.4 $\mathrm{M_{Jup}}$ for both solutions. In particular, we obtained the following orbital inclination solutions: 51.4$_{-11.5}^{+16.4}$° and 129.0$_{-16.2}^{+11.2}$°. Based on our analysis, we conclude that HD 6718 b is a planet. GDR3 reports a RUWE value of 0.957 for HD 6718 b. A significantly higher RUWE would be expected for a companion with a mass of $\sim$63.0 \mjup. The expected astrometric signal is limited, since the GDR3 astrometric baseline spans 1038 days, shorter than the planet’s 2476-day orbital period, and both periastron passages (in 2011 and 2018) occurred outside this time window.
As illustrated in Figure~\ref{fig:SC_HD6718b}, the companion mass derived in this work explains the current observed astrometric noise. Future astrometric monitoring covering a complete orbital period will enable a more precise determination of the companion’s true mass.

\subsection{HD 16760 b}
HD 16760 (HIP 12638) is a G5 star and it is likely part of a visual system with HD 16760 B (HIP 12635) \citep{abt1988maximum,sinachopoulos2007ccd}, which is 1.521 $\pm$ 0.002 mag fainter, separated by more than 700 au, and with an orbital period exceeding 10,000 years. HD 16760 hosts a companion first reported by \cite{sato2009substellar}. The authors analysed 27 RV data: ten obtained with the High Dispersion Spectrograph (HDS) on the Subaru Telescope \citep{noguchi2002high} from December 2004 to February 2008 and seventeen with the HIRES spectrograph on the Keck telescope between January 2006 and January 2009. They derived a $\mathrm{M_{c}}\sin{i}$ $\sim$ 13.1 \mjup, placing the companion at the transition between the planetary and brown dwarf regimes. After, \cite{kiefer2021determining} estimated an inclination of $\sim$3.2° and a mass of $\sim$219.9 \mjup, consistent with a M dwarf.

\begin{figure}
\centering
\includegraphics[width=.8\linewidth]{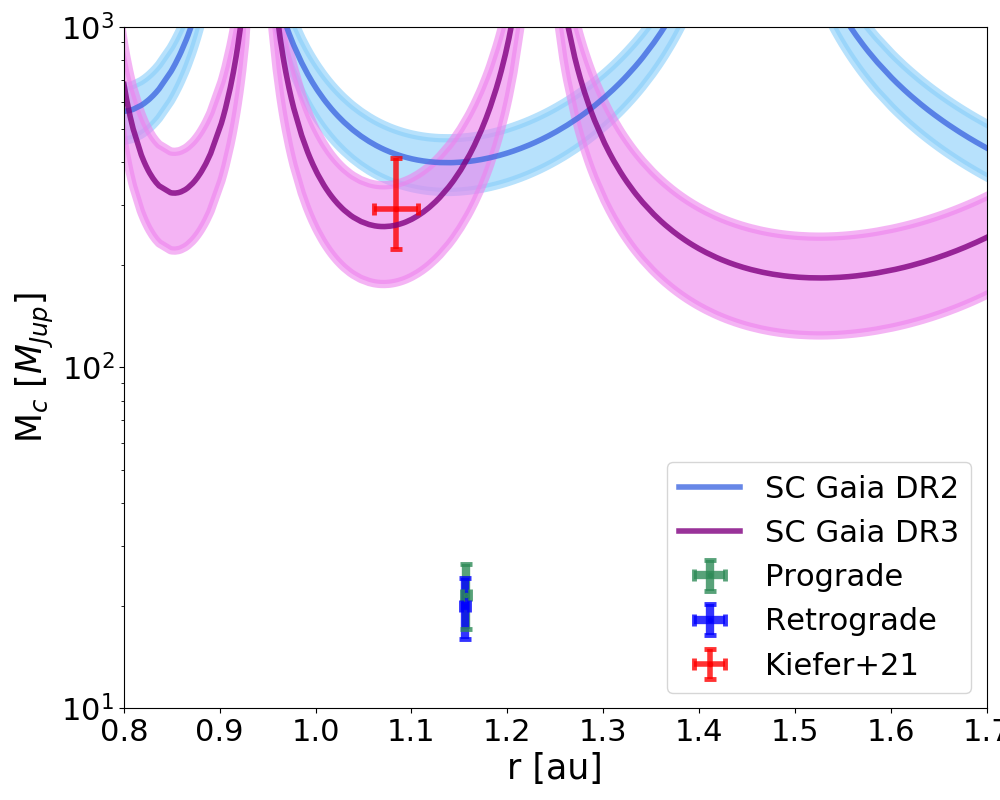}
\caption{Sensitivity Curve for HD 16760 companion. The red cross represents the value obtained by \cite{kiefer2021determining} using \texttt{GASTON}; the green cross is the prograde value and the blue cross is the retrograde value we obtained with GDR3 in this work: our results are not compatible with the sensitivity curve.} 
\label{fig:SC_HD16760b}
\end{figure}

\begin{figure}
\centering
\includegraphics[width=1\linewidth]{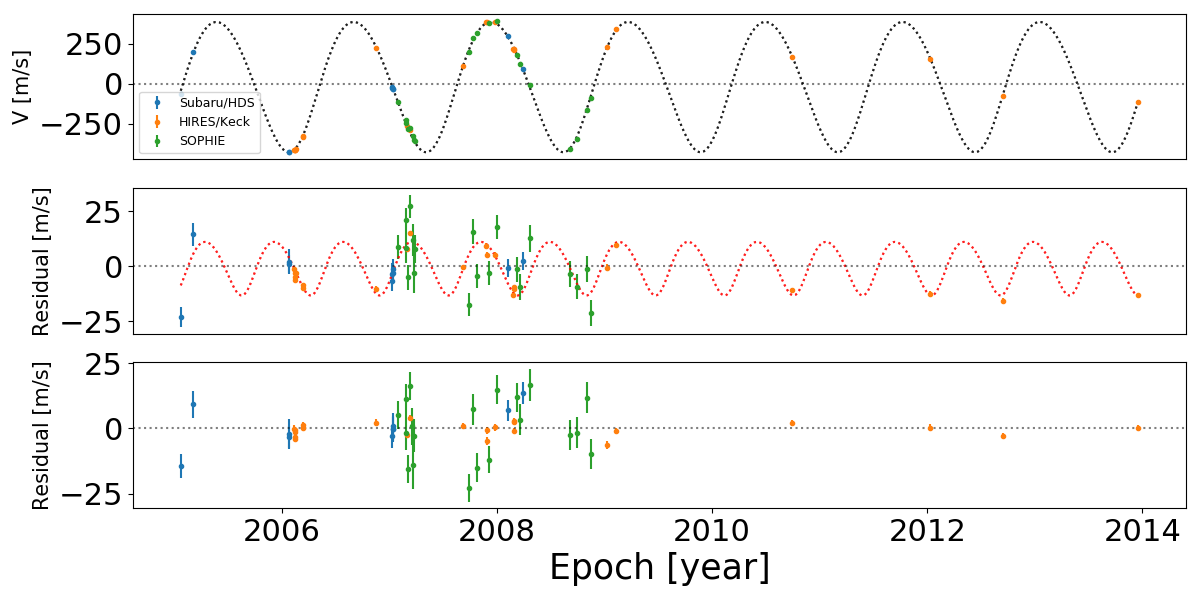}
\caption{\textit{Top:} RV data and Keplerian function of HD 16760 system. \textit{Middle:} Residual signal with the Keplerian function of the second object. \textit{Bottom:} Residuals of the fitted orbit of the second object.}
\label{fig:RV_HD16760b}
\end{figure}

In this study, we incorporated additional RV data obtained with SOPHIE/OHP \citep{Bouchy2006,perruchot2008sophie} from December 2006 to October 2008 collected by \cite{bouchy2009sophie} and HIRES/Keck data from \cite{2017AJ....153..208B}, for a total of 26 new RV data. We obtained mass values of 21.5$_{-4.5}^{+5.0}$ and 20.0$_{-4.0}^{+4.1}$ \mjup, for the prograde and retrograde configurations, where we defined the orbital inclination solutions as 43.3$_{-28.8}^{+32.2}$° and 132.6$_{-29.6}^{+30.3}$°, respectively. From our analysis, we classify HD 16760 b as a brown dwarf. We show the posterior distributions and the RV offsets and the jitter terms (Figure \ref{fig:CP_HD16760b}, Table \ref{tab:rvoffset_HD16760b}) in the Appendix. 
Our estimated masses lie below the sensitivity curve, suggesting the potential presence of an additional object (Figure \ref{fig:SC_HD16760b}). We analysed the residual signal in the RV fit using the GLS periodogram. This analysis revealed a signal with a semi-amplitude of 30.5 $\pm$ 2.6 ms$^{-1}$ and a period of 235.7 $\pm$ 1.5 days. 
We then studied the influence of the stellar activity, as the source of this signal. In \cite{bouchy2009sophie}, using SOPHIE data, the authors showed that the bisector span plotted against the RVs exhibited no significant variations (see Fig.2). Furthermore, we examined the S-index values, reported in \cite{2017AJ....153..208B} article, by plotting them as function of the RVs, which also displayed no correlation, indicating a low level of stellar activity. Additionally, the GLS analysis of the S-index data did not reveal any periodicity. 
The lack of a correlation with the RV data and the stellar activity indicators, along with the absence of any significant periodicity, strengthens our confidence that the detected signal originates from the host star's barycentric motion induced by an orbiting companion.
We performed a two planet-model fit in order to study the signal, adding a second Keplerian to our model. The RV data with the model of the first planet is displayed in the top plot in Figure \ref{fig:RV_HD16760b}; the middle plot shows a possible residual signal with a model for the second planet; the bottom plot shows the residuals. We evaluate a signal with \textsc{K} = 12.2$_{-4.3}^{+4.5}$ ms$^{-1}$, \textsc{P} = 234.3$_{-2.0}^{+2.5}$ days, e = 0.095$_{-0.063}^{+0.095}$, \textsc{$\omega$} = 207.2$_{-59.5}^{+58.4}$°, a = 0.729$_{-0.005}^{+0.006}$ au and a derived minimum mass of 0.4 $\pm$ 0.1 \mjup. Our semi-amplitude is lower than the one evaluate using GLS periodogram for two reasons: (1) we used a Keplerian model instead of a sinus function and (2) our model favoured the HIRES values since they have the lowest errors. Based on these information, we carefully say that we find a possible planet candidate. Future observations will gain a better constrictions on this signal. We employed the criteria presented in \cite{Trotta_BIC} based on the Bayesian Information Criterion (BIC) \citep{Schwarz1978,Kass1995} to evaluate which is the best model. The difference between the second and the first model is 9, marginally favouring the 1-planet model. 
Finally, the RUWE value for HD 16760 is 26.546, indicative of a significant astrometric noise, potentially due to HD 16760 b or another unseen companion. HD 16760 is presented in the Gaia DR3 non-single stars catalogue \citep{Holl2023,halbwachs2023}, reporting the period found for HD 16760 b and an orbital inclination of 9.407 $\pm$ 10.357°, which mildly favours a face-on configuration. As shown in Figure~\ref{fig:CP_HD16760b}, for our prograde solution, although the median inclination is 43.3$_{-28.8}^{+32.2}$°, the MAP (Maximum A Posteriori) value is $\sim$16.4°. This discrepancy arises from the different methods used to estimate the posterior distribution. Such low inclination implies a true mass of $\sim$52.4 $\mathrm{M_{Jup}}$, once again placing the companion within the BD regime. The presence of a BD, the probable presence of another object and another star make this target very interesting for the description of the formation and migration models. 


\subsection{30 Ari B b}
\begin{figure}
\centering
\includegraphics[width=.8\linewidth]{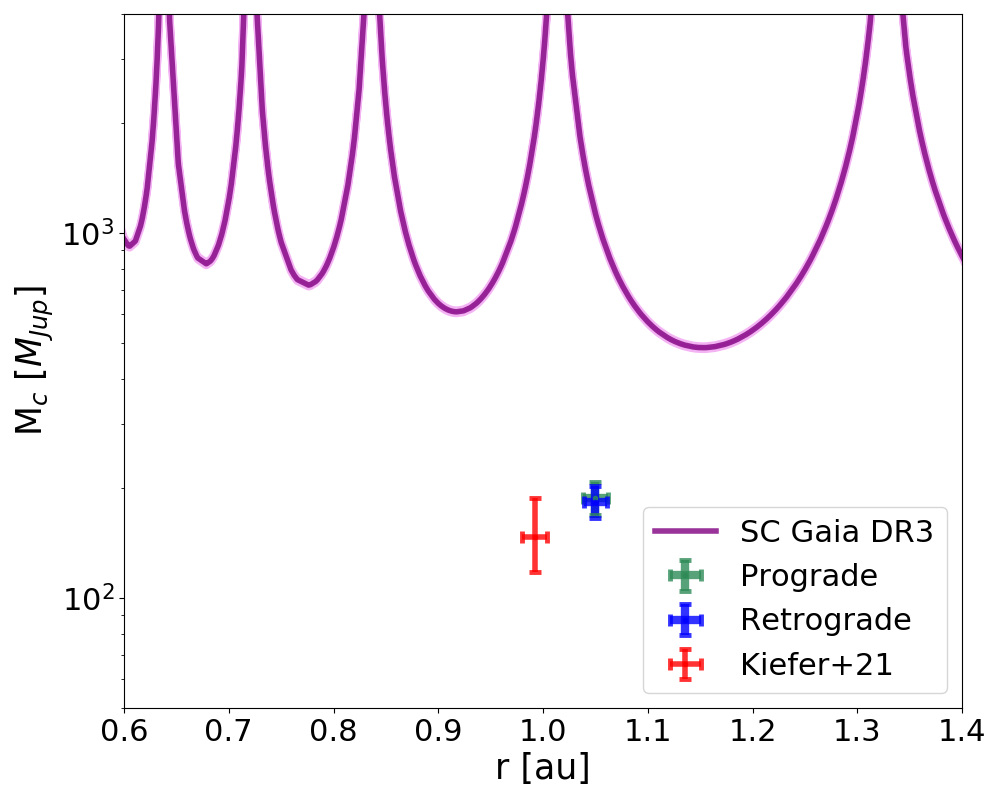}
\caption{Sensitivity Curve for 30 Ari B companion. The red cross represents the value obtained by \cite{kiefer2021determining} using \texttt{GASTON}; the green cross is the prograde value and the blue cross is the retrograde value we obtained with GDR3 in this work: our results are not compatible with the sensitivity curve.} 
\label{fig:SC_30AriBb}
\end{figure}

\begin{figure}
\centering
\includegraphics[width=1\linewidth]{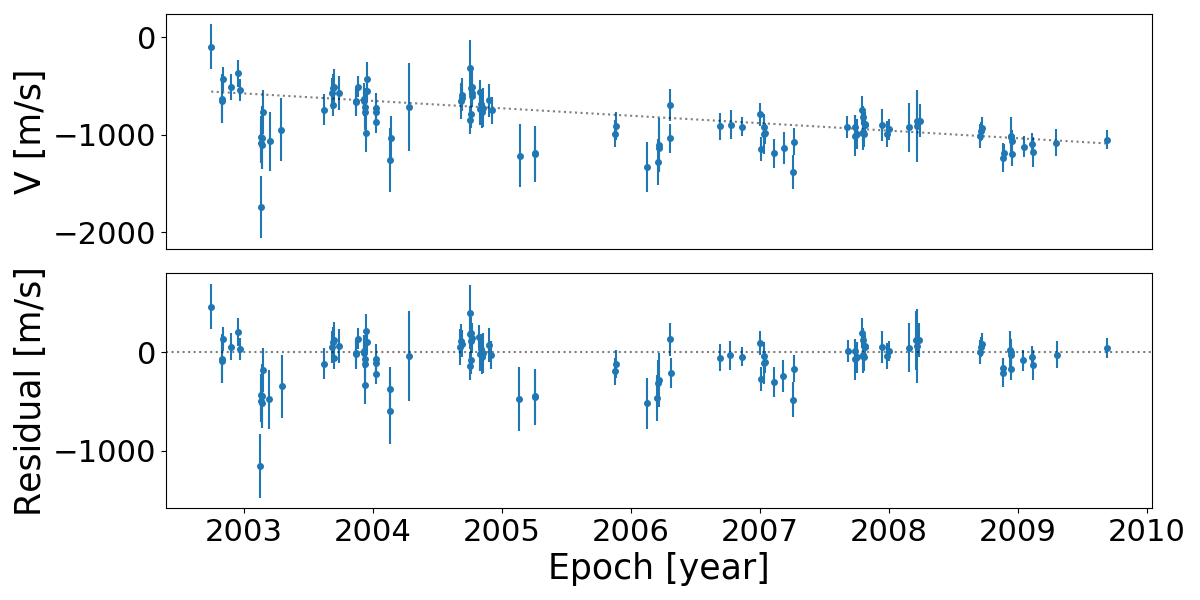}
\caption{\textit{Top:} RV data and Keplerian function of 30 Ari C. \textit{Bottom:} Residuals of the fitted orbit.} 
\label{fig:star_30AriBb}
\end{figure}

30 Ari B is a F6 star, part of a visual binary system: the second component 30 Ari A is a F5V star, with a projected distance between the B component of 1520 $\pm$ 54 au, as presented in \cite{guenther2009substellar}. In the same article, the discovery of 30 Ari B b was reported. The companion was identified with 98 RVs within the RV planet search program using the 2m-Alfred Jensch telescope of the Th$\mathrm{\ddot{u}}$ringer Landessternwarte Tautenburg \citep{hatzes2005}. \cite{guenther2009substellar} found a signal with \textsc{K} = 272 $\pm$ 24 \ms and a minimum mass of 9.88 $\pm$ 0.94 \mjup. \cite{2015ApJ...815...32K} added 12 RV points, which extend the time baseline $\sim$300 days, obtained with the same instrument and processed using the same pipeline, along with new photometric data, to better constrain the orbit, deriving \textsc{K} = 177 $\pm$ 26 \ms and a minimum mass of 6.6 $\pm$ 0.9 \mjup. 
They also identified a stellar companion, 30 Ari C, located at 22 au from 30 Ari B, estimating a mass of $>$ 0.29 $\mathrm{M_{\odot}}$. Also \cite{Roberts2015} reported the presence of 30 Ari C with a mass of 0.5 $\mathrm{M_{\odot}}$. \cite{kiefer2021determining} derived an orbital inclination of 4.181$_{-0.931}^{+1.031}$° and a corresponding true mass of 147.4$_{-29.5}^{+41.3}$ \mjup, for 30 Ari B b.
In our analysis, we utilized the same RV dataset used by \cite{2015ApJ...815...32K} and we considered only one keplerian to describe 30 Ari B b.
We found a true mass of 188.1$_{-18.5}^{+19.7}$ $\mathrm{M_{Jup}}$ with an orbital inclination of 2.9 $\pm$ 0.3°, for the prograde solution, and 184.0$_{-17.9}^{+18.6}$ $\mathrm{M_{Jup}}$ with 177.1 $\pm$ 0.3°, for the retrograde one. Both Kiefer et al. and our results classify the companion as an M-dwarf. Figure \ref{fig:CP_30AriBb} shows the posterior distributions and Table \ref{tab:rvoffset_30AriBb} reports the RV offset and the jitter term. Unfortunately, the semi-major axis of 30 Ari B b corresponds to a spike in the sensitivity curve, shown in Figure \ref{fig:SC_30AriBb} for both for GDR2 and GDR3, resulting in a loss of sensitivity. However, the plot shows that either Kiefer's results and ours remain below the curve. 
A RUWE value of 1.398 does not support the presence of a high-mass companion at 1 au, and the fact that 30 Ari B is not listed in the non-single stars catalogue reinforces the conclusion that 30 Ari C is the main source of the elevated PMA signal. We performed a fit to evaluate the effects of this stellar companion. We managed to recover the stellar signal, finding a mass of $0.51_{-0.09}^{+0.05}$ $\mathrm{M_{\odot}}$, a semi-major axis of $19.02_{-0.64}^{+1.53}$ au and an eccentricity of 0.71$_{-0.13}^{+0.15}$, corresponding to an orbital period of approximately 68.3 years. The mass we obtain is consistent with the value inferred from imaging by \cite{2015ApJ...815...32K}. In Figure \ref{fig:star_30AriBb}, we present the fitted model of 30 Ari C, which clearly shows a linear trend in the data. This high-mass body compromises the characterisation of the lower-mass companion, 30 Ari B b. Follow-up observations are necessary to fully characterize this complex system.

\subsection{HD 141937 b} 
\begin{figure}
\centering
\includegraphics[width=1\linewidth]{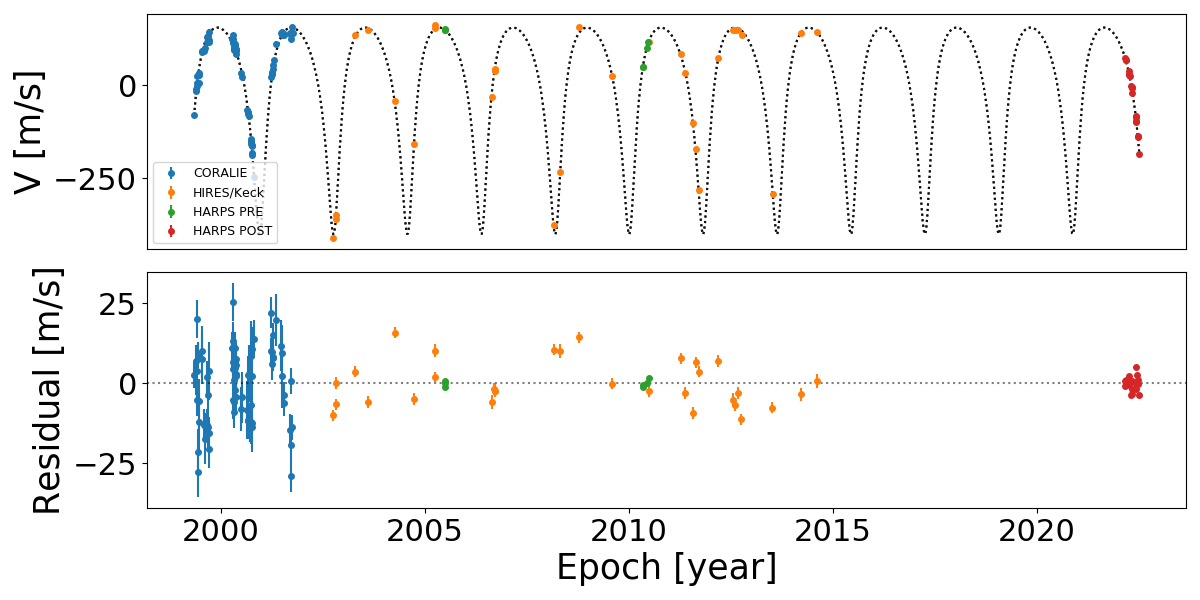}
\caption{\textit{Top:} RV data and Keplerian function of HD 141937 system. \textit{Bottom:} Residuals of the fitted orbit.}
\label{fig:RV_HD141937b}
\end{figure}

\begin{figure}
\centering
\includegraphics[width=.8\linewidth]{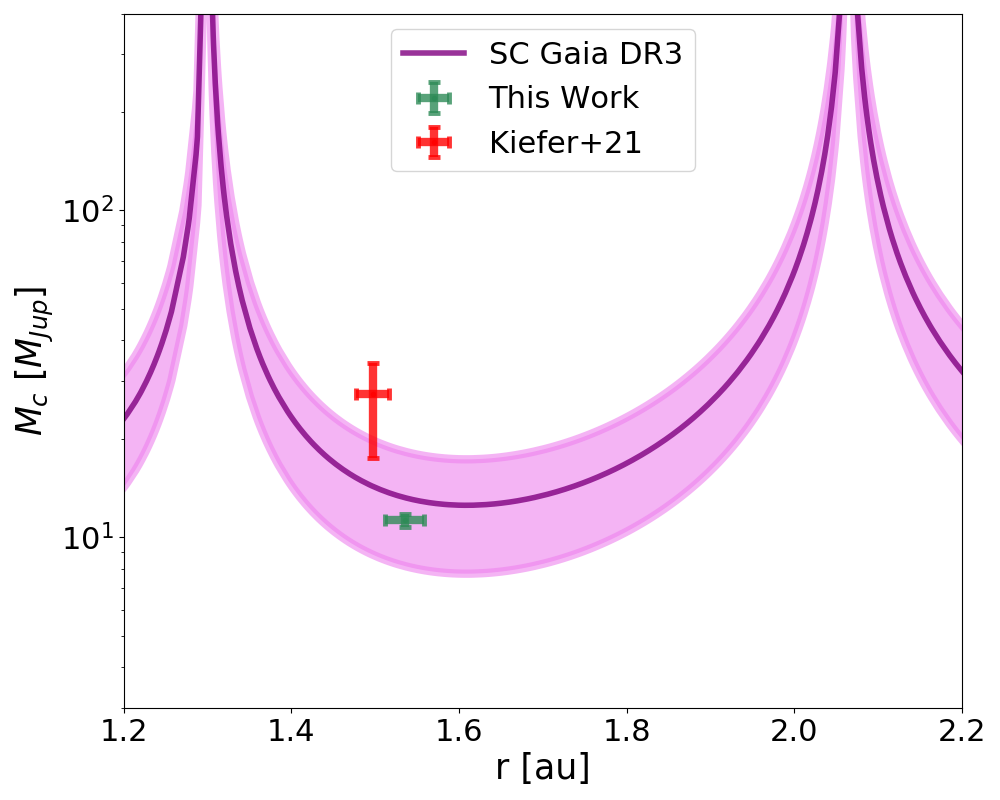}
\caption{Sensitivity Curve for HD 141937 companion. The red cross represents the value obtained by \cite{kiefer2021determining} using \texttt{GASTON}; the green cross is the value we obtained with GDR3 in this work: our result is compatible with the sensitivity curve.} 
\label{fig:SC_HD141937b}
\end{figure}

HD 141937 is a G1-type star hosting a substellar companion first reported by \citep{udry2002coralie}, based on 81 RV measurements obtained with CORALIE between March 1999 and August 2001.
Later, \cite{kiefer2021determining} estimated an inclination of $\sim$20.52° and a corresponding true mass of $\mathrm{M_{c}}$ = 27.4$_{-9.86}^{+6.78}$ M$\mathrm{_{Jup}}$, placing HD 141937 b within the BD regime at 1$\sigma$ confidence level, although a mass compatible with a planet could not be excluded within 2$\sigma$. 
In our analysis, we expanded the RV data by incorporating 57 additional points: thirty-one from HIRES/Keck (August 2002 to July 2014, \cite{2017AJ....153..208B}) obtained via the Data Analysis Center for Exoplanets (DACE) web platform, and twenty-six from HARPS (May 2005 to June 2022), divided into HARPS ``pre'' and HARPS ``post'' datasets based on instrumental upgrades, obtained from the ESO archive.
We performed separate analyses for prograde and retrograde solutions and noticed that the inclination tends towards an edge-on configuration. Consequently, we performed a comprehensive analysis over the full range of possible inclinations. This yielded an inclination of $90.00_{-6.76}^{+6.75}$° and a precise mass of 11.3 $\pm$ 0.5 \mjup, firmly classifying HD 141937 b as an edge-on giant planet. Figure \ref{fig:RV_HD141937b} shows the RV fit, the posterior distributions and the RV offsets with the jitter terms are reported in the Appendix (Fig. \ref{fig:CP_HD141937b}, Tab. \ref{tab:rvoffset_HD141937b}). The sensitivity curve (Fig. \ref{fig:SC_HD141937b}) supports our result, as the derived mass aligns with the expected astrometric signal. Additionally, we estimated the transit probability to be 0.32\%.
To verify the transit geometry, we calculated the angular range within which a transit could occur. Following \cite{transitprob}, we computed the opening angle $\theta$, which represents the maximum angular deviation between our line of sight and the observed inclination that still allows for a planetary transit. For an eccentric Keplerian orbit, $\theta$ is defined as:
\begin{equation}
\centering
    \theta = \arcsin{\left(\frac{R_{\star}}{a}\frac{1 + e\sin{\omega}}{1 - e^{2}}\right)}.
    \label{theta}
\end{equation}
For HD 141937 b, we obtained $\theta \sim$ 0.20°. This implies that the planet could transit if its orbital inclination lies within 89.80° $\leq$ \textit{i} $\leq$ 90.20°. In our analysis, we derived an inclination of \textit{i} $\sim$ 90.00° and we conclude that, considering the uncertainties, HD 141937 b could indeed be a transiting planet.
TESS observed this object in Sector 91. However, we do not detect any transit, which could be due to two main reasons: (1) the object has an orbital period of 652 days, while the TESS observation window is only about 27 days, making it unlikely to capture a transit event; and (2) the orbital inclination we derived has large uncertainties ($\pm$6.8°) and given that the transit geometry allows for only a narrow opening angle of about 0.20°, the object could easily fall outside the range that permits transits to be visible.
Recently, \cite{HD141937Wallace} utilized the RUWE to constrain the properties of potential companions. By simulating the expected astrometric signals, they aimed to refine the masses and orbital parameters of planet candidates detected with other methods, under the assumption that elevated RUWE values are primarily attributable to the presence of a companion. HD 141937, exhibiting a notably high RUWE value of 4.972, was among the first targets tested with this approach. Their analysis yielded a companion mass of $\mathrm{M_{c}}$ = 23.5$_{-5.1}^{+4.5}$ M$_{\mathrm{Jup}}$ and an inclination of 153 $\pm$ 2°. However, they cautioned that the reported result should be considered as an upper limit. 
Although the high RUWE value, HD 141937 does not appear in the Gaia DR3 non-single stars catalogue, indicating a poor astrometric fit: this system will be particularly interesting to study with Gaia DR4 data.

\subsection{HD 148427 b}
\begin{figure}
\centering
\includegraphics[width=.8\linewidth]{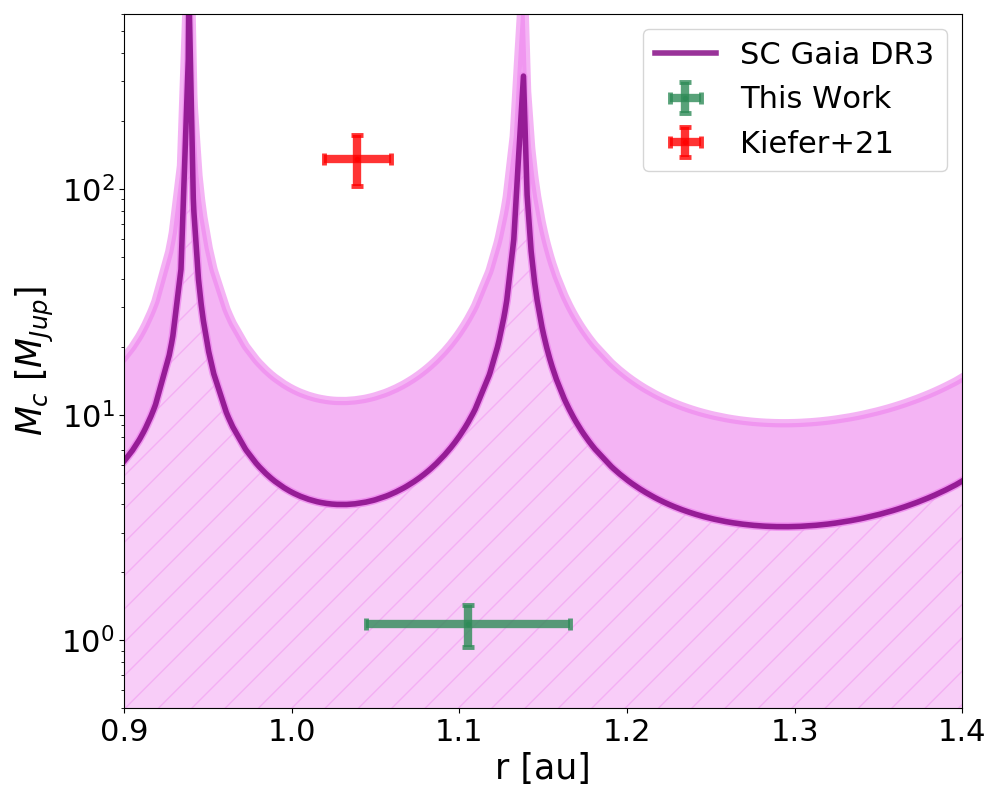}
\caption{Sensitivity Curve for HD 148427 companion. The red cross represents the value obtained by \cite{kiefer2021determining} using \texttt{GASTON}; the green cross is the value we obtained with GDR3 in this work: our result is below the sensitivity curve. Since the error of $\Delta\nu_{T}$ is bigger than the value itself, for construction, we can not represent it, because it would show a negative mass. With the hatched area we represent the positive values.} 
\label{fig:SC_HD148427b}
\end{figure}

HD 148427 is K0 subgiant star. It hosts a companion discovered by \cite{Fischer2009HD148427b} using 31 RV points obtained with Hamilton/Lick \citep{LickHamilton}, collected between July 2001 and February 2009. Later, Kiefer et al. (\citeyear{kiefer2021determining}) determined an orbital inclination of $\sim$0.51° and a corresponding mass of $\sim$136.5 \mjup, suggesting a stellar classification for the companion.
In our analysis, we utilized the same RV dataset and derived an inclination of 90.2$_{-30.7}^{+30.5}$°. Due to the  high inclination, our results differ from those of \cite{kiefer2021determining}, we obtained a lower companion mass of 1.2$_{-0.3}^{+0.2}$ \mjup. In Table \ref{tab:rvoffset_HD148427b}, we reported the RV offset and the jitter terms. 
The posterior distributions plot (Fig. \ref{fig:CP_HD148427b}) illustrates that some parameters are not tightly constrained, in particular the inclination. We identified a plausible range for the inclination of 60°-120°, so we cannot say that the mass we obtained is the true mass.
Nevertheless, we plot the sensitivity curve (Fig. \ref{fig:SC_HD148427b}) and the mass value obtained remains below the curve. 
The RUWE value of 1.055 suggests consistency with a substellar object, but it is too low to support the presence of a stellar companion.

\subsection{HD 96127 b} 
\begin{figure}
\centering
\includegraphics[width=.8\linewidth]{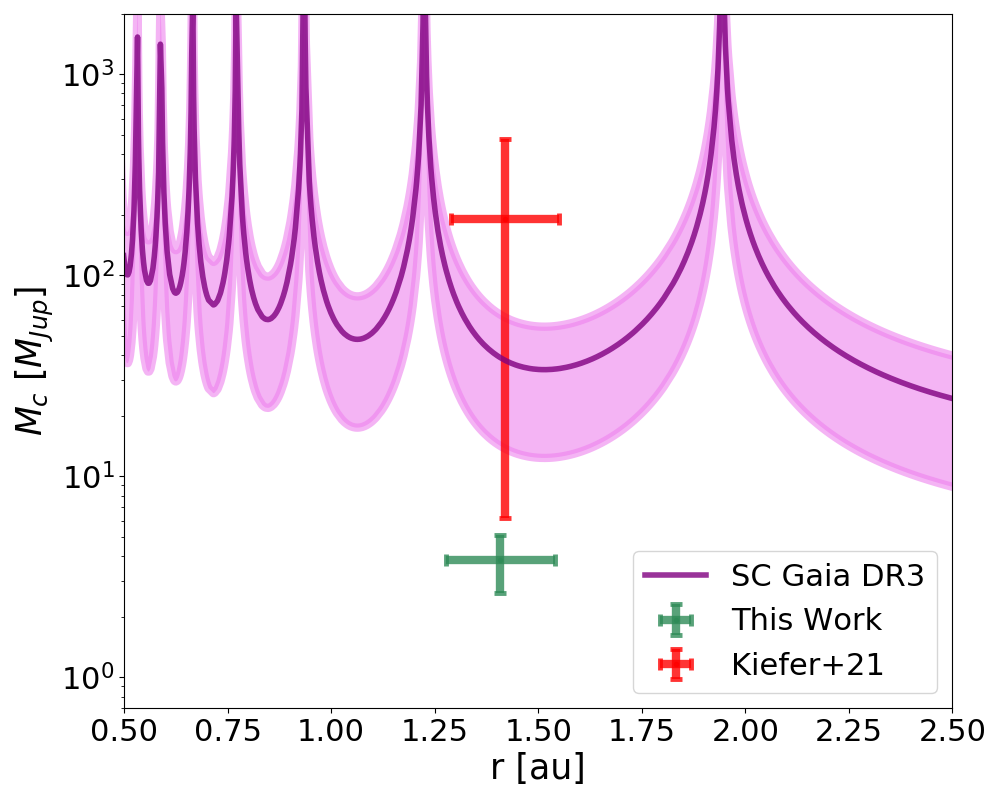}
\caption{Sensitivity Curve for HD 96127 companion. The red cross represents the value obtained by \cite{kiefer2021determining} using \texttt{GASTON}; the green cross is the value we obtained with GDR3 in this work: our result is below the sensitivity curve.} 
\label{fig:SC_HD96127b}
\end{figure}

HD 96127 is a K2 giant star, hosting a companion discovered by \cite{gettel2012hd96127b} using 50 RV points obtained with the HRS/HET between January 2004 and February 2009.
Kiefer et al. (\citeyear{kiefer2021determining}) estimated an \textit{i} = 1.4$_{-0.8}^{+38.5}$° and $\mathrm{M_{c}}$ = 190$_{-184}^{+284}$ \mjup, placing the object within the stellar regime. However, the authors noted convergence issues.
In our analysis, we reanalysed the same RV points. The posterior distributions and the RV offset and the jitter term are reported in the Appendix (Fig. \ref{fig:CP_HD96127b}), Tab. \ref{tab:rvoffset_HD96127b}). As reported also in \cite{gettel2012hd96127b}, the jitter term has a large value of $\sim$50 $\mathrm{ms^{-1}}$. We derived a mass of $\sim$3.8 \mjup, and we constrained the inclination within a wide range between 50°-130°. The sensitivity curve (Fig. \ref{fig:SC_HD96127b}) shows that our derived mass falls below the curve. This could be due to a bad constriction of the parameters or the presence of an additional companion. Indeed, \cite{gettel2012hd96127b} identified a 25-days periodicity in the \Hipparcos photometry time series, suggesting a potential second companion that warrants further investigation with new observations. We also performed a GLS analysis on the residuals: we do not found this periodicity, but a 6-days signal with analytical FAP $>$ 0.01 \%. More photometric observations and RV data are needed to identify the nature of these signals. The RUWE value of 0.997 is consistent with the companion mass we derived, while does not support the presence of a stellar companion.

\subsection{HIP 65891 b}
HIP 65891 is a giant K0 star, one of the most massive that are known to host a substellar companion. \cite{jones2015hip65891} used 49 RV data points to analyse HIP 65891 b: twenty-six collected by CHIRON/Cerro Tololo Inter-American Observatory \citep{chiron} and twenty-three collected with FEROS/La Silla observatory \citep{feros}. Kiefer et al. (\citeyear{kiefer2021determining}) derived an orbital inclination of 1.184$_{-0.207}^{+0.256}$° and a corresponding true mass of 312.3$_{-57.4}^{+74.2}$ \mjup, classifying the companion as an M-dwarf.
In our analysis, we re-examined the same RV dataset, and similarly to HD 148427 b, we obtained a high inclination within the range 50°-130°, that bring to a low mass of 6.1 $\pm$ 1.3 $\mathrm{M_{Jup}}$.
We initially performed the analysis using GDR3. Given that the orbital period closely matches to the GDR3 observational window (1038 days), it corresponds to one of the sensitivity ``spike'' of the curve, where sensitivity precision is significantly reduced. For this reason, we re-analysed the system using GDR2 data. 

\begin{figure}
\centering
\includegraphics[width=.8\linewidth]{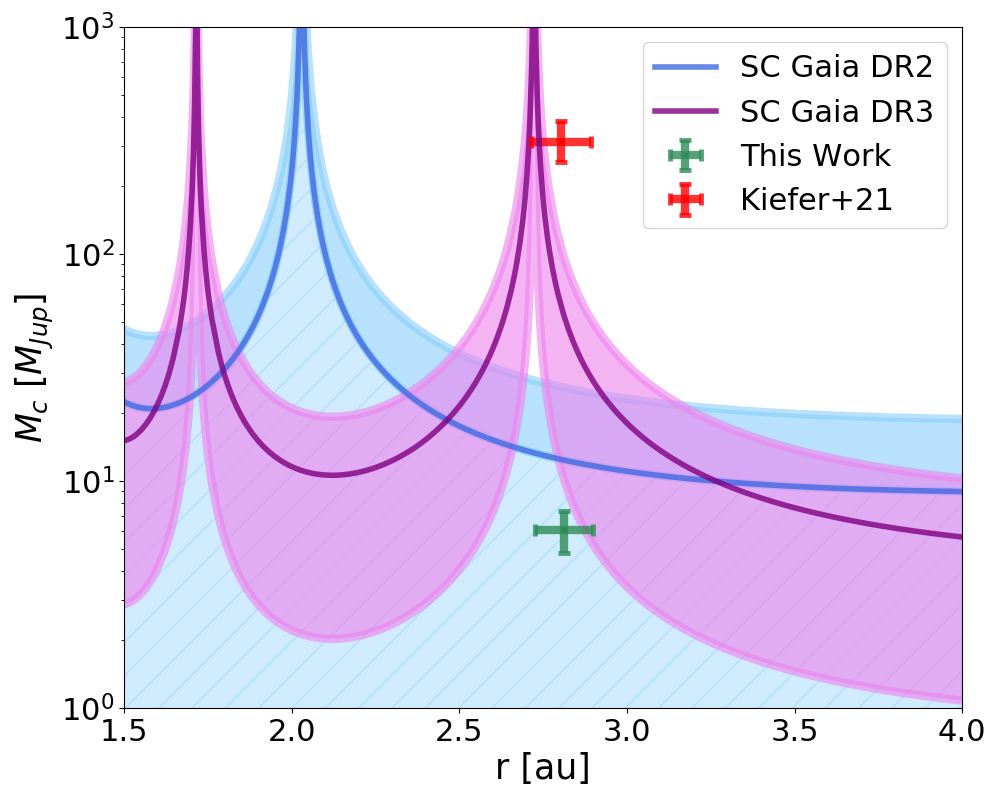}
\caption{Sensitivity Curve for HIP 65891 companion. The red cross represents the value obtained by \cite{kiefer2021determining} using \texttt{GASTON}; the green cross is the value we obtained with GDR2 in this work: our result is compatible with the sensitivity curve.} 
\label{fig:SC_HIP65891b}
\end{figure}

The posterior distributions and the RV offsets and the jitter terms are reported in the Appendix (Fig. \ref{fig:CP_HIP65891b}, Tab. \ref{tab:rvoffset_HIP65891b}). The sensitivity curve (Fig. \ref{fig:SC_HIP65891b}) indicates that the derived mass is consistent within 1$\sigma$ with the GDR2 sensitivity curve, supporting our result with the astrometric signal measured. Additionally, HIP 65891 has a RUWE value of 0.904, which is lower than what would typically be expected if the companion were an M-dwarf. This value supports our findings.

\section{Conclusions}
\label{sec:conclusion}

We presented a method to determining the true mass of long-period planets by combining RV data with the PMA technique. To assess the reliability of our results, we compared them with the sensitivity curve, a theoretical tool described in \cite{kervella2019stellar}. As a validation, we applied our model to three benchmark systems (GJ 463 b, $\pi$ Men b and HD 222237 b). We subsequently applied, for the first time, a joint RV and astrometric analysis to a small sample of objects previously studied with GDR1, with the aim of refining their mass estimates using the latest data from GDR3.

We studied HD 5388 b and HD 6718 b that are formerly classified as BDs/low-mass stars. Our analysis indicates more probable masses of $\sim$3.2 $\mathrm{M_{Jup}}$ and $\sim$2.4 \mjup, respectively. These values suggest that both objects are likely planets, in agreement with the sensitivity curve. In particular, HD 5388 exhibits a RUWE value that is too low to support the classification of HD 5388 b as a BD/M-dwarf star, contrary to interpretations reported in the existing literature. Furthermore, we identified an additional signal in the RV residuals that may be indicative of another sub-stellar companion. Using all available RV data of HD 6718 system, we were able to constrain the mass of HD 6718 b, which is in agreement with the current astrometric noise. In this case as well, the RUWE value supports a planetary classification.
\cite{kiefer2021determining} described HD 16760 b as a M-dwarf star, with a mass consistent with the astrometric signal indicated by the sensitivity curve. However, our analysis yields a mass in the BD regime ($21.5_{-4.5}^{+5.0}$ $\mathrm{M_{Jup}}$) and identifies a further signal that may correspond to a planetary-mass companion.

Our derived mass of 30 Ari B b, falling within the stellar regime, is consistent with the value reported by \cite{kiefer2021determining}. Several authors have suggested the presence of a stellar companion (30 Ari C), and we also managed to recover its signal. This companion appears to have a significant impact on the PMA measurements, contrary to the internal companion, leading to an inaccurate characterisation of the system.

HD 141937 b is currently classified as a BD, with reported orbital inclinations of approximately 20.5° and 153°, for the two solutions. In contrast, our analysis, based on the inclusion of nearly 60 publicly available HARPS RV data, yields a mass within the planetary regime, consistent with the sensitivity curve, and an orbital inclination compatible with a quasi-transiting configuration.
For HD 148427 b, HD 96127 b and HIP 65891 b, we did not manage to constrain firmly the orbital inclinations. HD 148427 b and HD 96127 b are, at the time of writing, classified as stellar companion. From our analysis we found a plausible range for \textit{i} equal to 60°-120° and a mass of $\sim$1.2 $\mathrm{M_{Jup}}$ and a range 50°-130° and a mass of $\sim$3.8 \mjup, respectively. Although both mass estimates lie below the sensitivity curve, suggesting that the derived masses cannot fully explain the detected astrometric signal, their RUWE values remain consistent with a substellar object. HIP 65891 b is currently classified as a M-dwarf. Using GDR2, we found the orbital inclination within 50°-130° range and a mass of $\sim$6.1 \mjup: this value is slightly below the sensitivity curve. As in the previous cases, the RUWE value for HIP 65891 is too low to support a stellar nature for the companion.

The masses derived in this work are, in almost all cases, lower than those obtained through simulations with \texttt{GASTON}. This discrepancy likely arises from the methodology employed by Kiefer et al., where orbital inclination, and consequently mass, was constrained using simulated astrometric excess noise. 
The GDR1 astrometric data do not account for effects caused by orbital motion in binary systems. As a consequence, astrometric excess noise absorbs modelling errors from unresolved binaries, along with instrumental modelling errors \citep{GDR12016}. Moreover, as highlighted by \citet{2021A&A...651A..11D}, the \texttt{GASTON} tool does not model multiple companions in multi-object systems. When only one component is considered, the derived mass can be overestimated if other companions significantly contribute to the astrometric signal. Finally, it is important to note that the time baseline of GDR1 (418 days) is shorter than the orbital period of most of the targets in our sample, further limiting the reliability of excess noise as a robust constraint on orbital inclination. Finally, the authors of \texttt{GASTON} have recently refined their analysis method through the development of \texttt{GaiaPMEX}, which benefits from the higher-quality data provided by GDR3 \citep{Kiefer2024}.

There are already different models to recover the true masses of stellar companions by combining RV and astrometric data. In most cases, our code allows us to place tighter constraints on the system parameters. We also incorporate the sensitivity curve—completely independent of our model—as a key element in evaluating the robustness of the obtained results. 
Unlike as other integrated RV-astrometric models, we have already implemented an additional feature to account for stellar activity using Gaussian Processes. 
Our goal is to further integrate our model with the sensitivity curve in order to study multiplanetary systems and to determine the inclinations and true masses of hidden companions. In doing so, we aim to gain new information on systems hosting long-period planets and to expand our understanding of their formation and dynamical evolution.
\begin{acknowledgements}
We are truly grateful to the referee Timothy Brandt for the relevance of his comments that greatly improved the quality of the paper.
G.P., A.P., G.M. acknowledge support from the ASI-INAF agreement 2021-5-HH.2-2024 and the European Union - Next Generation EU through the grant n. 2022J7ZFRA - Exo-planetary Cloudy Atmospheres and Stellar High energy (Exo-CASH) funded by MUR - PRIN 2022. Based on data obtained from the ESO Science Archive Facility with DOI: \url{https://doi.org/10.18727/archive/33}. This work has made use of data from the European Space Agency (ESA) mission {\it Gaia} (\url{https://www.cosmos.esa.int/gaia}), processed by the {\it Gaia} Data Processing and Analysis Consortium (DPAC, \url{https://www.cosmos.esa.int/web/gaia/dpac/consortium}). Funding for the DPAC has been provided by national institutions, in particular the institutions participating in the {\it Gaia} Multilateral Agreement. This publication makes use of The Data \& Analysis Center for Exoplanets (DACE), which is a facility based at the University of Geneva (CH) dedicated to extrasolar planets data visualisation, exchange and analysis. DACE is a platform of the Swiss National Centre of Competence in Research (NCCR) PlanetS, federating the Swiss expertise in Exoplanet research. The DACE platform is available at https://dace.unige.ch.
\end{acknowledgements}
\bibliographystyle{aa} 
\bibliography{bibliography}

\begin{appendix}
\label{appendix}

\section{RV offset and jitter terms}
In our analysis, we derived the RV offset and the jitter term for every instrument and dataset considered for each target. In this section we report the results for all the systems analysed.
\begin{table}[h!]
\small
 \caption{RV offsets and jitter terms of GJ 463 system.}
    \centering
    \renewcommand\arraystretch{1.2}
    \begin{tabular}{lc}
    \hline
    \hline
         Parameter & Value \\
         \hline
         $\gamma$ [ms$^{-1}$] (HRS/HET) & 12.0 $_{-2.0}^{+2.1}$ \\
         $\gamma$ [ms$^{-1}$] (HIRES/Keck) & 10.6$_{-2.6}^{+2.5}$ \\
         $\sigma$ [ms$^{-1}$] (HRS/HET) &  0.001$_{-0.005}^{+0.348}$ \\
         $\sigma$ [ms$^{-1}$] (HIRES/Keck) & 6.5$_{-1.9}^{+2.3}$ \\
        \hline
        \hline
             \end{tabular}
    \label{tab:rvoffset_gj463b}
\end{table}

\begin{table}[h!]
\small
 \caption{RV offsets and jitter terms of $\pi$ Men system.}
    \centering
    \renewcommand\arraystretch{1.2}
    \begin{tabular}{lc}
    \hline
    \hline
         Parameter & Value \\
         \hline
         $\gamma$ [ms$^{-1}$] (UCLES/AAT) & 2.0 $\pm$ 1.0 \\
         $\gamma$ [ms$^{-1}$] (CORALIE 98) & 10671.4 $\pm$ 4.1 \\
         $\gamma$ [ms$^{-1}$] (CORALIE 07) & 10674.4 $\pm$ 4.1 \\
         $\gamma$ [ms$^{-1}$] (CORALIE 14) & 10699.3 $\pm$ 0.9 \\
         $\gamma$ [ms$^{-1}$] (HARPS ``pre'') & 10709.3 $\pm$ 0.3 \\
         $\gamma$ [ms$^{-1}$] (HARPS ``post'') & 10731.3 $\pm$ 0.5 \\
         $\sigma$ [ms$^{-1}$] (UCLES/AAT) & 3.8$_{-0.9}^{+1.0}$ \\
         $\sigma$ [ms$^{-1}$] (CORALIE 98) & 9.5$_{-2.6}^{+4.4}$ \\
         $\sigma$ [ms$^{-1}$] (CORALIE 07) & 12.0$_{-2.7}^{+3.8}$ \\
         $\sigma$ [ms$^{-1}$] (CORALIE 14) & 2.7$_{-0.8}^{+0.9}$ \\
         $\sigma$ [ms$^{-1}$] (HARPS ``pre'') & 2.7 $\pm$ 0.2 \\
         $\sigma$ [ms$^{-1}$] (HARPS ``post'') & 1.6$_{-0.3}^{+0.4}$  \\
        \hline
        \hline
             \end{tabular}
    \label{tab:rvoffset_piMenb}
\end{table}

\begin{table}[h!]
\small
 \caption{RV offsets and jitter terms of HD 222237 system.}
    \centering
    \renewcommand\arraystretch{1.2}
    \begin{tabular}{lc}
    \hline
    \hline
         Parameter & Value \\
         \hline
         $\gamma$ [ms$^{-1}$] (PSF/Magellan II) & -24.4$_{-2.4}^{+2.6}$ \\
         $\gamma$ [ms$^{-1}$] (UCLES/AAT) & 17.7 $\pm$ 2.1 \\
         $\gamma$ [ms$^{-1}$] (HARPS ``pre'') & 69862.7$_{-2.4}^{+2.5}$ \\
         $\gamma$ [ms$^{-1}$] (HARPS ``post'') & 69873.8$_{-2.8}^{+2.9}$ \\
         $\sigma$ [ms$^{-1}$] (PSF/Magellan II) & 2.0 $\pm$ 0.2 \\
         $\sigma$ [ms$^{-1}$] (UCLES/AAT) & 4.4$_{-0.7}^{+0.8}$ \\
         $\sigma$ [ms$^{-1}$] (HARPS ``pre'') & 1.9$_{-0.2}^{+0.3}$ \\
         $\sigma$ [ms$^{-1}$] (HARPS ``post'') & 5.2$_{-1.0}^{+1.5}$ \\
        \hline
        \hline
             \end{tabular}
    \label{tab:rvoffset_HD222237b}
\end{table}

\begin{table}[h!]
\small
 \caption{RV offsets and jitter terms of HD 5388 system.}
    \centering
    \renewcommand\arraystretch{1.2}
    \begin{tabular}{lc}
    \hline
    \hline
         Parameter & Value \\
         \hline
         $\gamma$ [ms$^{-1}$] (HARPS ``pre'') & 39315.4$_{-0.7}^{+0.8}$ \\
         $\gamma$ [ms$^{-1}$] (HARPS ``post'') & 39331.3 $\pm$ 1.4 \\
         $\sigma$ [ms$^{-1}$] (HARPS ``pre'') & 2.6$_{-0.4}^{+0.5}$\\
         $\sigma$ [ms$^{-1}$] (HARPS ``post'') & 0.031$_{-0.008}^{+0.750}$ \\
        \hline
        \hline
             \end{tabular}
    \label{tab:rvoffset_HD5388b}
\end{table}

\begin{table}[h!]
\small
 \caption{RV offsets and jitter terms of HD 6718 system.}
    \centering
    \renewcommand\arraystretch{1.2}
    \begin{tabular}{lc}
    \hline
    \hline
         Parameter & Value \\
         \hline
         $\gamma$ [ms$^{-1}$] (HARPS ``pre'') & 34750.8 $\pm$ 0.8 \\
         $\gamma$ [ms$^{-1}$] (HARPS ``post'') & 34766.1$_{-4.1}^{+3.8}$ \\
         $\sigma$ [ms$^{-1}$] (HARPS ``pre'') & 1.7$_{-0.3}^{+0.4}$ \\
         $\sigma$ [ms$^{-1}$] (HARPS ``post'') & 4.8$_{-2.2}^{+4.3}$ \\
        \hline
        \hline
             \end{tabular}
    \label{tab:rvoffset_HD6718b}
\end{table}

\begin{table}[h!]
\small
 \caption{RV offsets and jitter terms of HD 16760 system.}
    \centering
    \renewcommand\arraystretch{1.2}
    \begin{tabular}{lc}
    \hline
    \hline
         Parameter & Value \\
         \hline
         $\gamma$ [ms$^{-1}$] (HDS/Subaru) & -0.435 $\pm$ 1.9 \\
         $\gamma$ [ms$^{-1}$] (HIRES/Keck) & -109.1 $\pm$ 0.7 \\
         $\gamma$ [ms$^{-1}$] (SOPHIE) & -3560.0 $\pm$ 2.7 \\
         $\sigma$ [ms$^{-1}$] (HDS/Subaru) & 3.7$_{-2.6}^{+4.1}$ \\
         $\sigma$ [ms$^{-1}$] (HIRES/Keck) & 2.6$_{-0.5}^{+0.7}$ \\
         $\sigma$ [ms$^{-1}$] (SOPHIE) & 9.5$_{-2.0}^{+2.5}$ \\
        \hline
        \hline
             \end{tabular}
    \label{tab:rvoffset_HD16760b}
\end{table}

\begin{table}[h!]
\small
 \caption{RV offsets and jitter terms of 30 Ari B system.}
    \centering
    \renewcommand\arraystretch{1.2}
    \begin{tabular}{lc}
    \hline
    \hline
         Parameter & Value \\
         \hline
         $\gamma$ [ms$^{-1}$] (Alfred Jensch) & -21.2 $\pm$ 15.4\\
         $\sigma$ [ms$^{-1}$] (Alfred Jensch) & 39.3$_{-0.05}^{+16.06}$ \\
        \hline
        \hline
             \end{tabular}
    \label{tab:rvoffset_30AriBb}
\end{table}

\begin{table}[h!]
\small
 \caption{RV offsets and jitter terms of HD 141937 system.}
    \centering
    \renewcommand\arraystretch{1.2}
    \begin{tabular}{lc}
    \hline
    \hline
         Parameter & Value \\
         \hline
         $\gamma$ [ms$^{-1}$] (CORALIE) & -2926.4 $\pm$ 1.4 \\
         $\gamma$ [ms$^{-1}$] (HIRES/Keck) & -39.2 $\pm$ 1.4 \\
         $\gamma$ [ms$^{-1}$] (HARPS ``pre'') & -2880.2 $\pm$ 1.0 \\
         $\gamma$ [ms$^{-1}$] (HARPS ``post'') & -2868.1 $\pm$ 1.0 \\
         $\sigma$ [ms$^{-1}$] (CORALIE) & 9.7$_{-1.0}^{+1.1}$ \\
         $\sigma$ [ms$^{-1}$] (HIRES/Keck) & 7.1$_{-1.0}^{+1.2}$ \\
         $\sigma$ [ms$^{-1}$] (HARPS ``pre'') & 0.8$_{-0.4}^{+0.6}$ \\
         $\sigma$ [ms$^{-1}$] (HARPS ``post'') & 2.2$_{-0.4}^{+0.5}$ \\
        \hline
        \hline
             \end{tabular}
    \label{tab:rvoffset_HD141937b}
\end{table}

\begin{table}[h!]
\small
 \caption{RV offsets and jitter terms of HD 148427 system.}
    \centering
    \renewcommand\arraystretch{1.2}
    \begin{tabular}{lc}
    \hline
    \hline
         Parameter & Value \\
         \hline
         $\gamma$ [ms$^{-1}$] (Hamilton/Lick) & -0.4$\pm$ 1.6\\
         $\sigma$ [ms$^{-1}$] (Hamilton/Kick) & 4.9$_{-1.3}^{+1.5}$ \\
        \hline
        \hline
             \end{tabular}
    \label{tab:rvoffset_HD148427b}
\end{table}

\begin{table}[h!]
\small
 \caption{RV offsets and jitter terms of HD 96127 system.}
    \centering
    \renewcommand\arraystretch{1.2}
    \begin{tabular}{lc}
    \hline
    \hline
         Parameter & Value \\
         \hline
         $\gamma$ [ms$^{-1}$] (HRS/HET) & -917.6$_{-8.5}^{+8.2}$ \\
         $\sigma$ [ms$^{-1}$] (HRS/HET) & 48.0$_{-4.9}^{+5.9}$ \\
        \hline
        \hline
             \end{tabular}
    \label{tab:rvoffset_HD96127b}
\end{table}

\begin{table}[h!]
\small
 \caption{RV offsets and jitter terms of HIP 65891 system.}
    \centering
    \renewcommand\arraystretch{1.2}
    \begin{tabular}{lc}
    \hline
    \hline
         Parameter & Value \\
         \hline
         $\gamma$ [ms$^{-1}$] (FEROS) & 3.9 $\pm$ 3.0\\
         $\gamma$ [ms$^{-1}$] (CHIRON) & 7.7$_{-4.8}^{+4.6}$ \\
         $\sigma$ [ms$^{-1}$] (FEROS) & 8.2$_{-2.0}^{+2.4}$ \\
         $\sigma$ [ms$^{-1}$] (CHIRON) & 7.1$_{-1.5}^{+1.8}$ \\
        \hline
        \hline
             \end{tabular}
    \label{tab:rvoffset_HIP65891b}
\end{table}

\section{Results of the systems}
In this section we present the posterior distributions of all analysed systems. 

\begin{figure}
\centering
\includegraphics[width=1\linewidth]{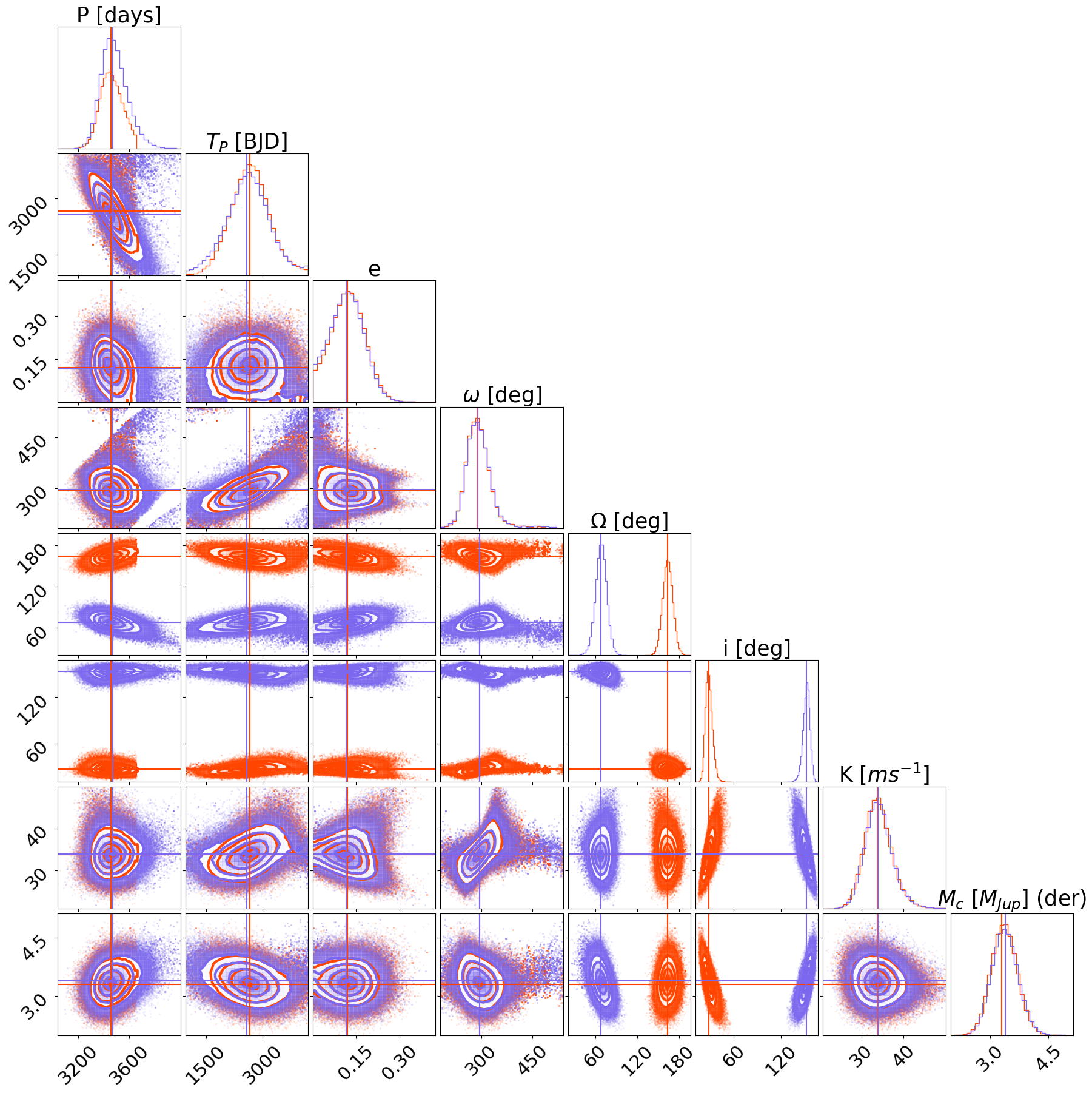}
\caption{Posterior distributions of GJ 463 system for the prograde (orange) and the retrograde (blue) solution.}
\label{fig:CP_gj463b}
\end{figure}

\begin{figure}
\centering
\includegraphics[width=1\linewidth]{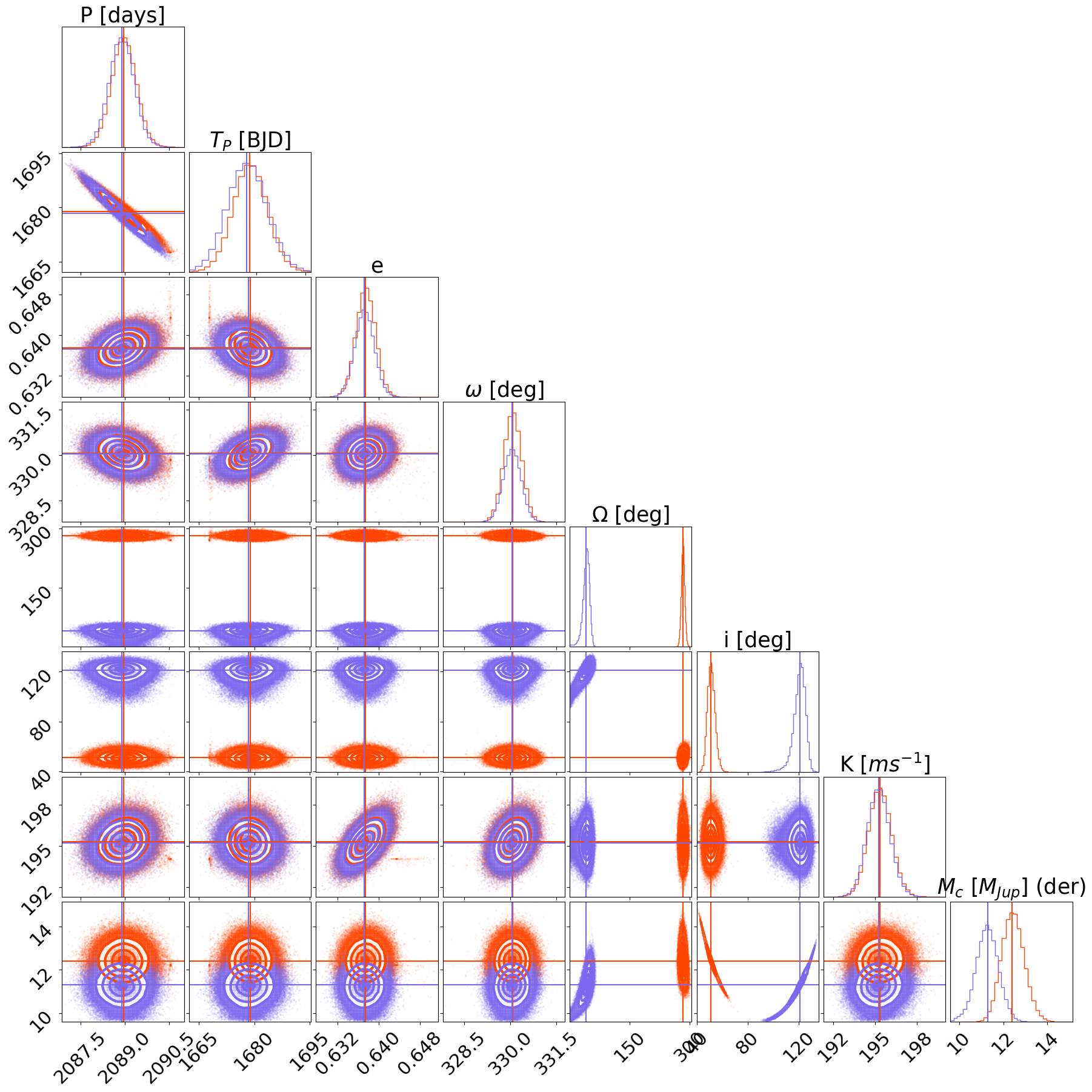}
\caption{Posterior distributions of $\pi$ Men system.}
\label{fig:CP_piMenb}
\end{figure}

\begin{figure}
\centering
\includegraphics[width=1\linewidth]{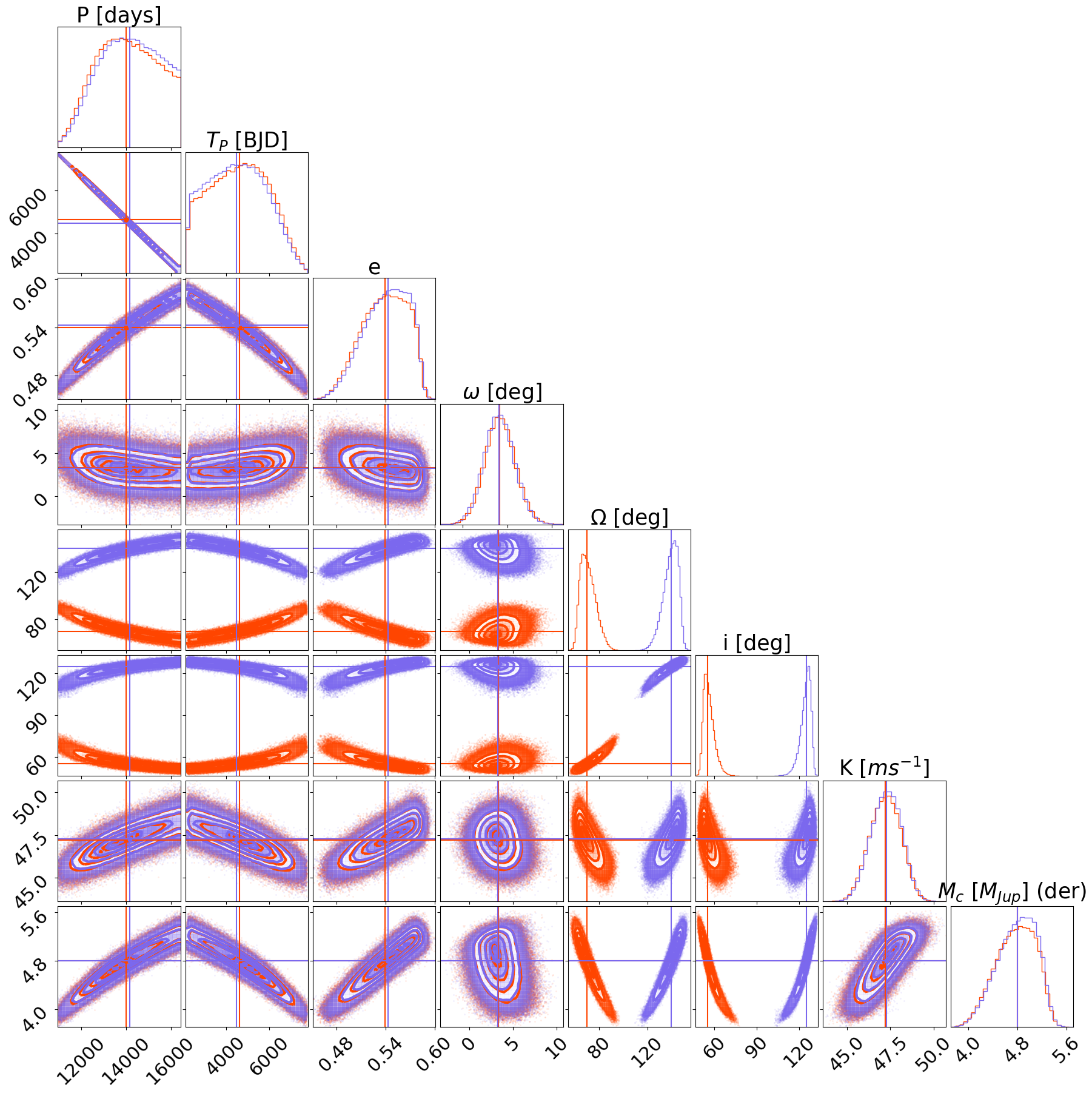}
\caption{Posterior distributions of HD 222237 system.}
\label{fig:CP_HD222237b}
\end{figure}

\begin{figure}
\centering
\includegraphics[width=1\linewidth]{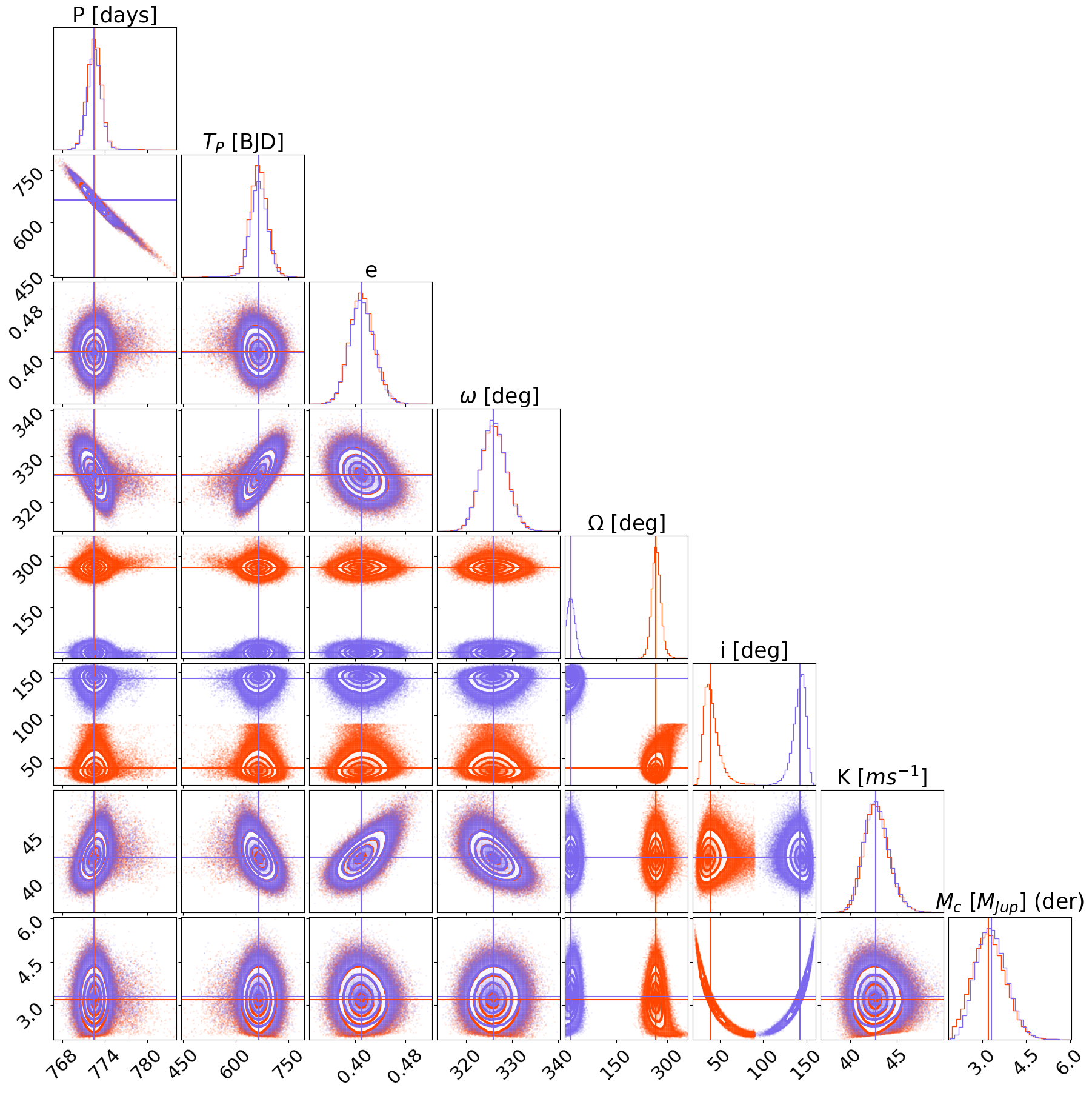}
\caption{Posterior distributions of HD 5388 system.}
\label{fig:CP_HD5388b}
\end{figure}

\begin{figure}
\centering
\includegraphics[width=1\linewidth]{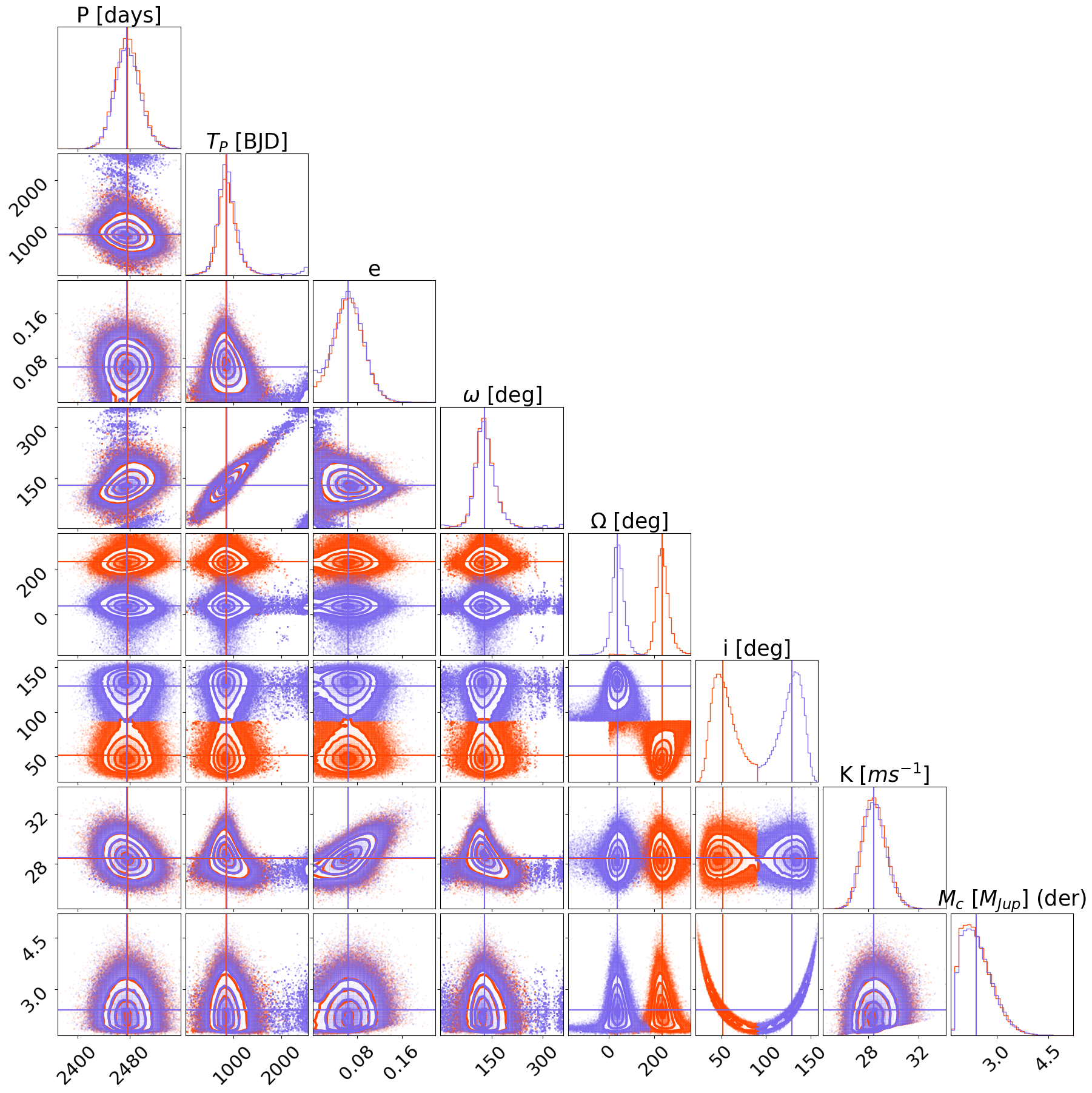}
\caption{Posterior distributions of HD 6718 system.}
\label{fig:CP_HD6718b}
\end{figure}

\begin{figure}
\centering
\includegraphics[width=1\linewidth]{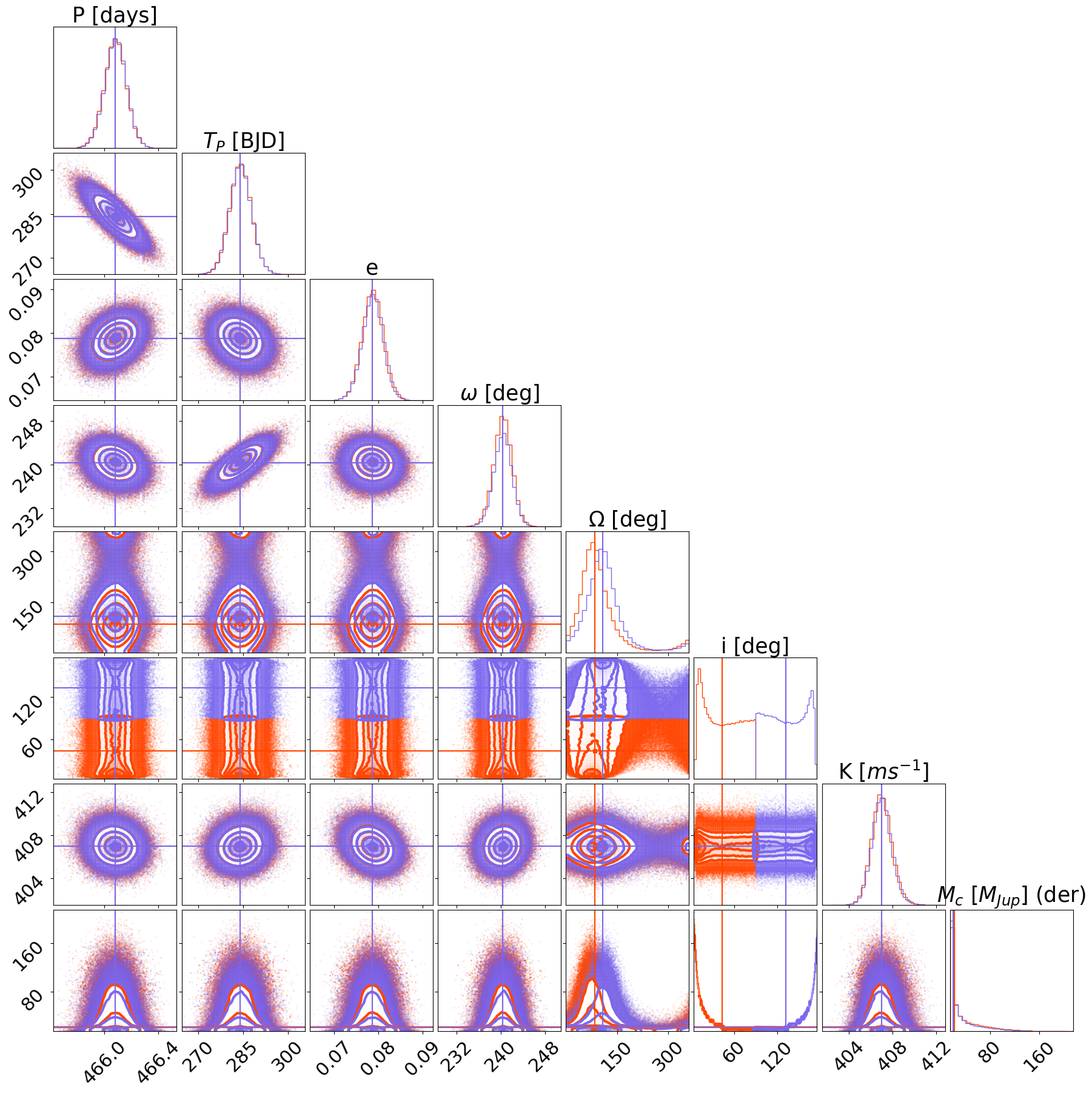}
\caption{Posterior distributions of HD 16760 system.}
\label{fig:CP_HD16760b}
\end{figure}

\begin{figure}
\centering
\includegraphics[width=1\linewidth]{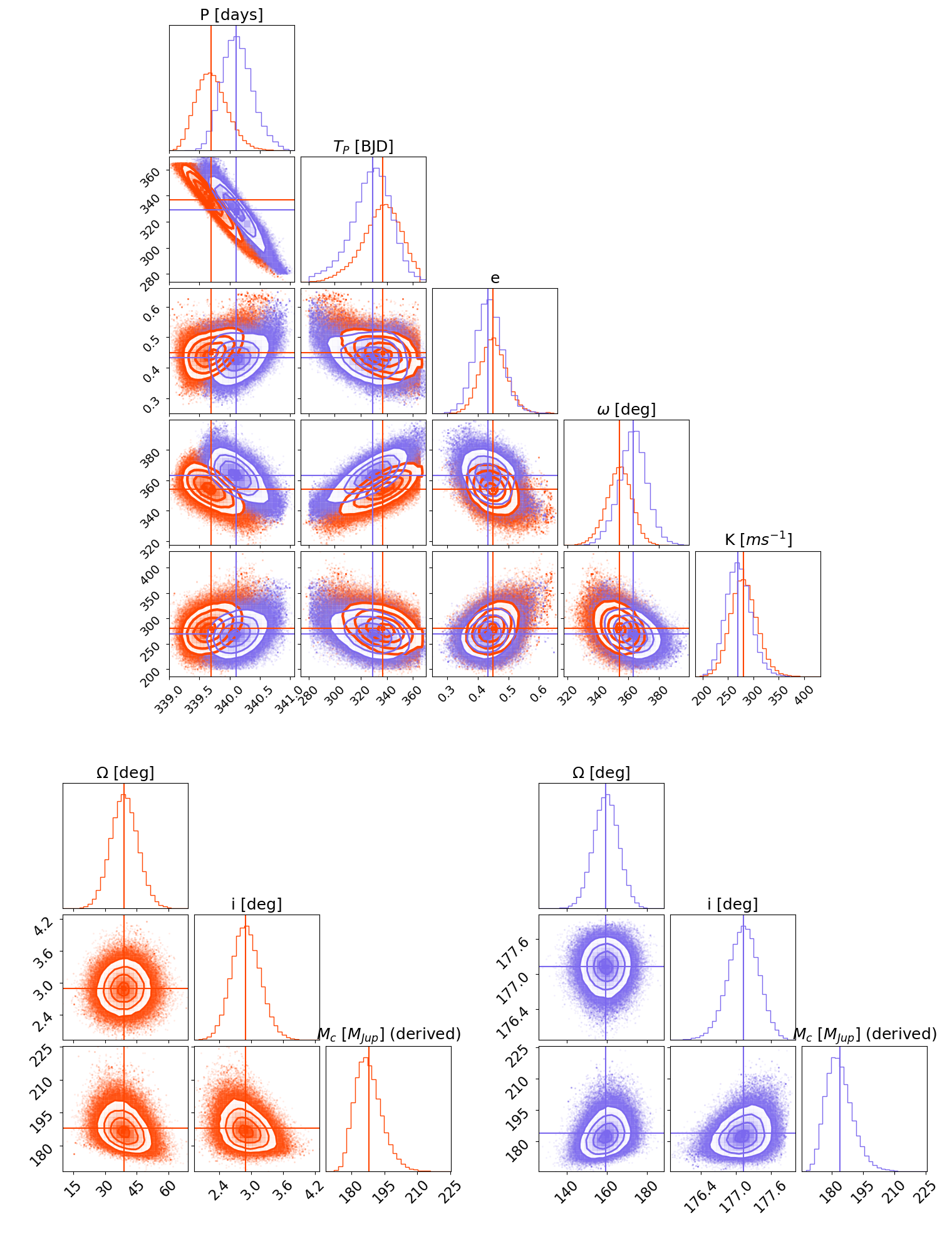}
\caption{Posterior distributions of 30 Ari B system. Astrometric and spectroscopic parameters are presented separately for clarity.}
\label{fig:CP_30AriBb}
\end{figure}

\begin{figure}
\centering
\includegraphics[width=1\linewidth]{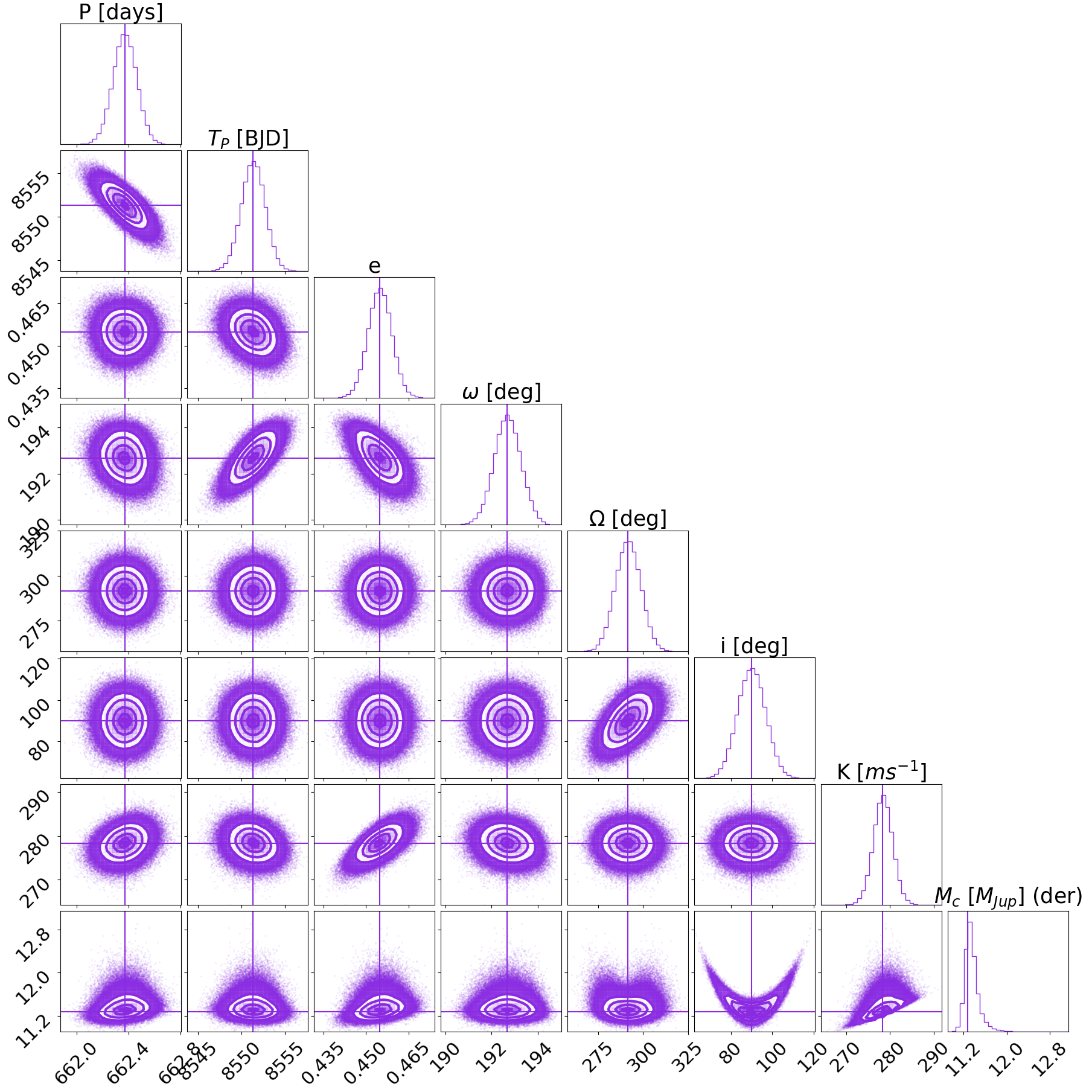}
\caption{Posterior distributions of HD 141937 system over the full range of orbital inclinations.}
\label{fig:CP_HD141937b}
\end{figure}

\begin{figure}
\centering
\includegraphics[width=1\linewidth]{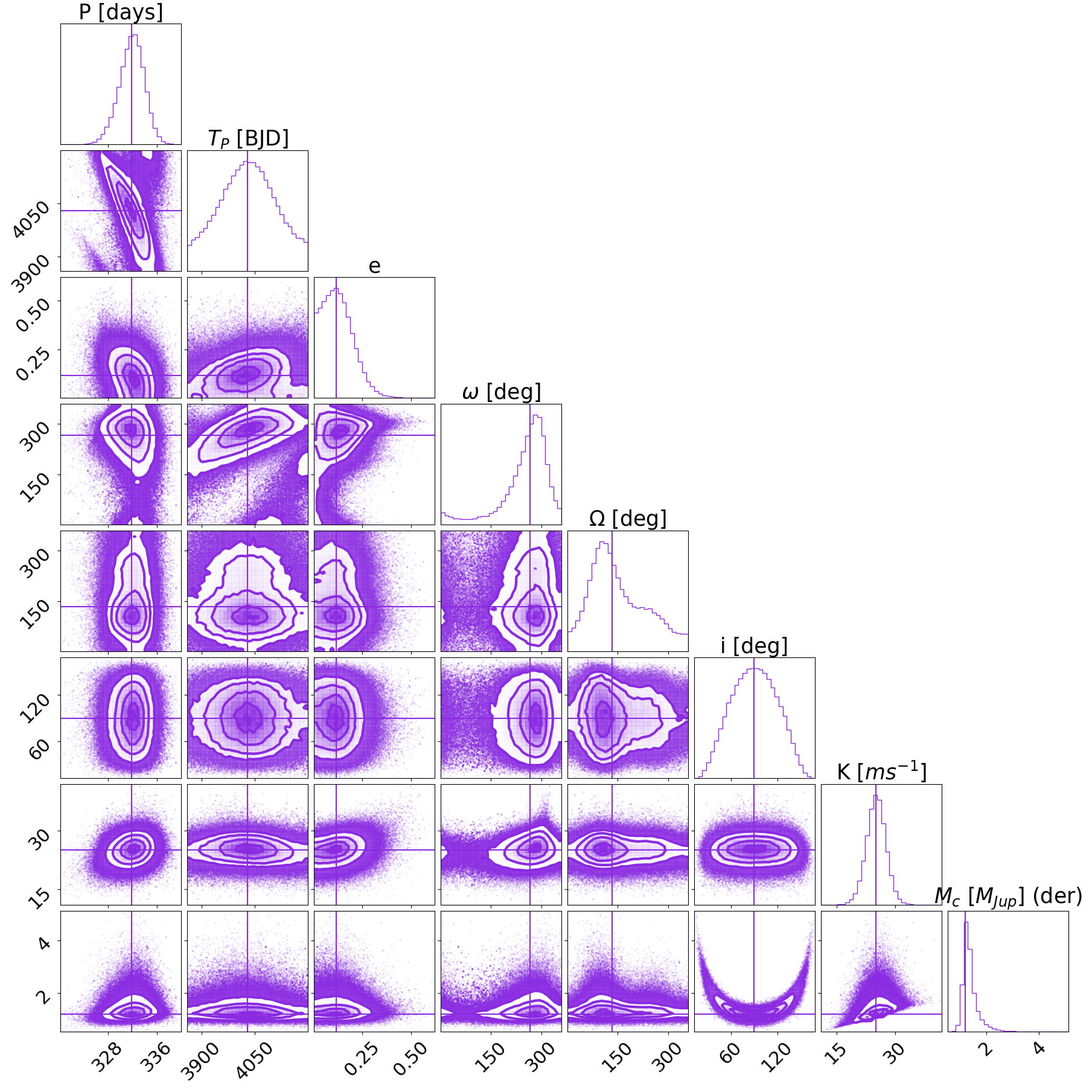}
\caption{Posterior distributions of HD 148427 system.}
\label{fig:CP_HD148427b}
\end{figure}

\begin{figure}
\centering
\includegraphics[width=1\linewidth]{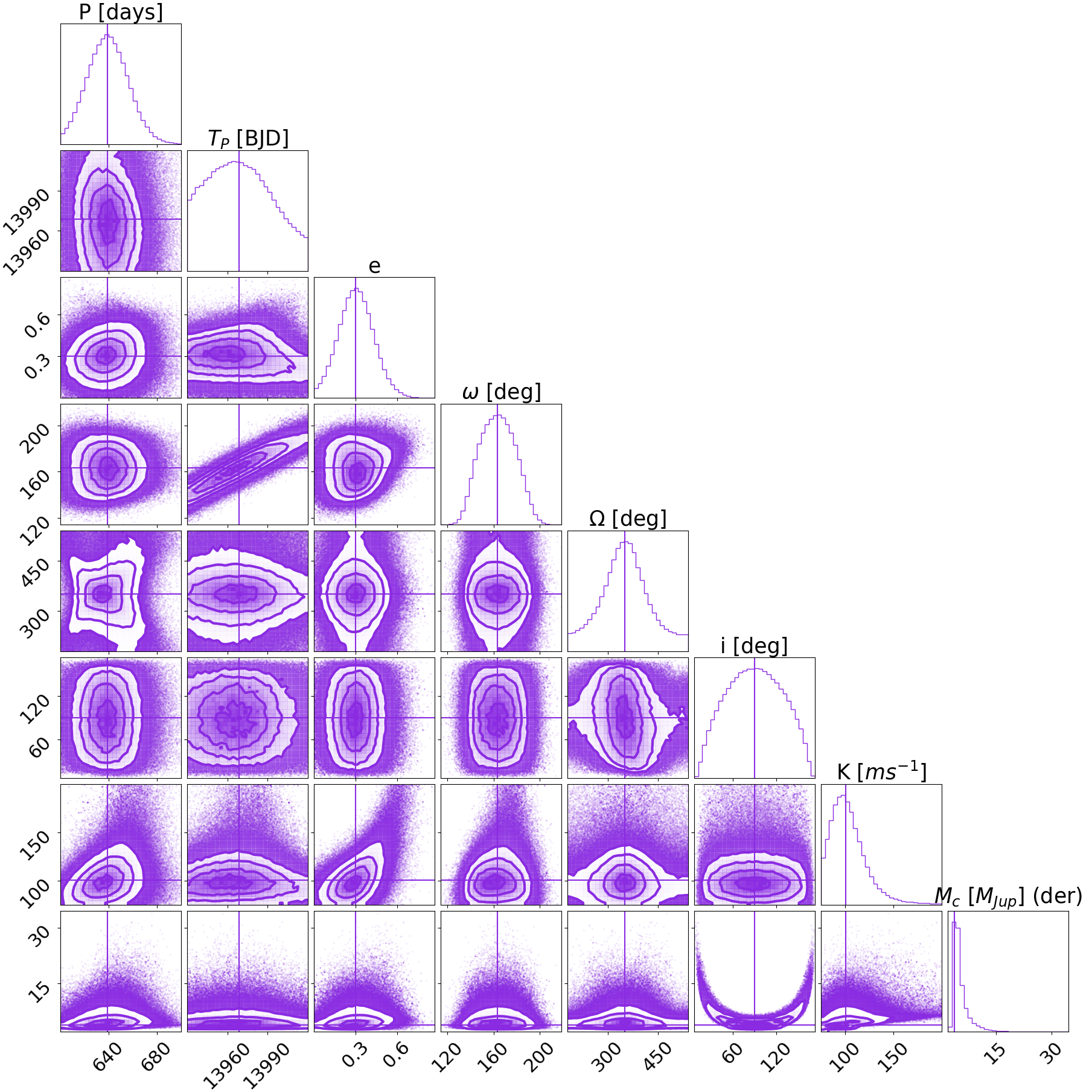}
\caption{Posterior distributions of HD 96127 system.}
\label{fig:CP_HD96127b}
\end{figure}

\begin{figure}
\centering
\includegraphics[width=1\linewidth]{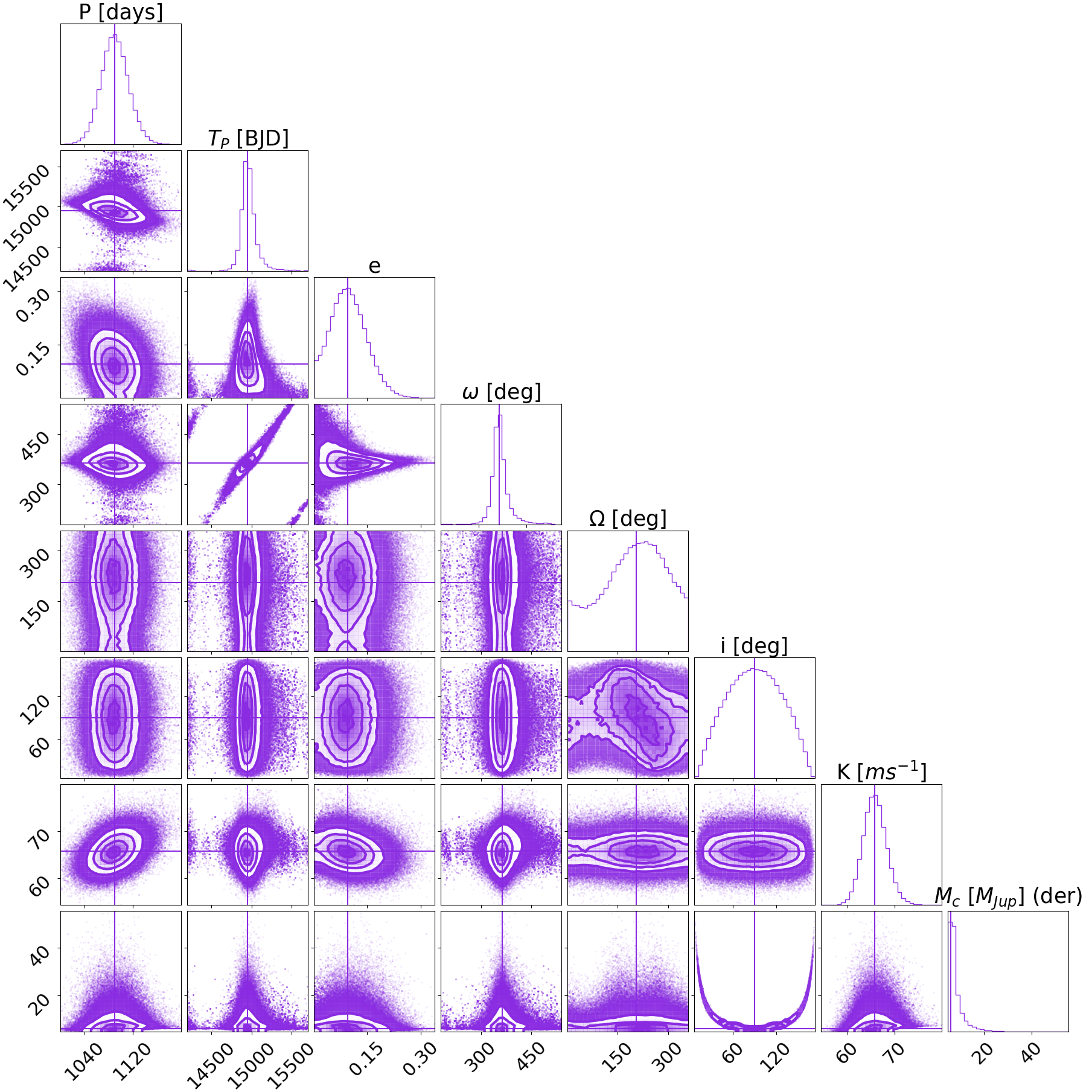}
\caption{Posterior distributions of HIP 65891 system.}
\label{fig:CP_HIP65891b}
\end{figure}

\end{appendix}

\end{document}